\documentclass[a4paper,twocolumn,11pt,accepted=2025-01-28]{quantumarticle}
\pdfoutput=1
\usepackage{tikz}
\usepackage{lipsum}
\usepackage[T1]{fontenc}
\usepackage[utf8]{inputenc}
\usepackage{float}
\usepackage{amsmath}
\usepackage{amssymb}
\usepackage{graphicx}
\usepackage{xcolor}
\usepackage{ulem}
\usepackage{latexsym}
\usepackage{epstopdf}
\usepackage{parskip}
\usepackage[english]{babel}
\usepackage{comment}
\usepackage{notes2bib}
\usepackage{indentfirst}
\usepackage{hyperref}
\usepackage[square,sort&compress,numbers]{natbib}
\makeatletter
\@ifundefined{textcolor}{}
{%
	\definecolor{BLACK}{gray}{0}
	\definecolor{WHITE}{gray}{1}
	\definecolor{RED}{rgb}{1,0,0}
	\definecolor{GREEN}{rgb}{0,1,0}
	\definecolor{BLUE}{rgb}{0,0,1}
	\definecolor{CYAN}{cmyk}{1,0,0,0}
	\definecolor{MAGENTA}{cmyk}{0,1,0,0}
	\definecolor{YELLOW}{cmyk}{0,0,1,0}
}

\makeatother

\begin{document}
		
\title{Light-matter correlations in Quantum Floquet engineering of cavity quantum materials}

\author{Beatriz Pérez-González}
\email{perezg.bea@gmail.com}
\affiliation{Instituto de Ciencia de Materiales de Madrid (ICMM), CSIC, Calle Sor Juana Inés de la Cruz 3, 28049 Madrid, Spain}
\affiliation{Institute of Physics, University of Augsburg, Augsburg, 86159, Germany }
\author{Gloria Platero}
\email{gplatero@icmm.csic.es}
\affiliation{Instituto de Ciencia de Materiales de Madrid (ICMM), CSIC, Calle Sor Juana Inés de la Cruz 3, 28049 Madrid, Spain}
\author{Álvaro Gómez-León}
\email{a.gomez.leon@csic.es}
\affiliation{Instituto de F\'isica Fundamental (IFF), CSIC, Calle Serrano 113b, 28006 Madrid, Spain.}


	\begin{abstract}
    Quantum Floquet engineering (QFE) seeks to generalize the control of quantum systems with classical external fields, widely known as Semi-Classical Floquet engineering (SCFE), to quantum fields. However, to faithfully capture the physics at arbitrary coupling, a gauge-invariant description of light-matter interaction in cavity-QED materials is required, which makes the Hamiltonian highly non-linear in photonic operators.
    We provide a non-perturbative truncation scheme of the Hamiltonian, which is valid or arbitrary coupling strength, and use it to investigate the role of light-matter correlations, which are absent in SCFE.
    We find that even in the high-frequency regime, light-matter correlations can be crucial, in particular for the topological properties of a system.
    As an example, we show that for a SSH chain coupled to a cavity, light-matter correlations break the original chiral symmetry of the chain, strongly affecting the robustness of its edge states.
    In addition, we show how light-matter correlations are imprinted in the photonic spectral function and discuss their relation with the topology of the bands.
	\end{abstract}

\maketitle

\section{Introduction}
Harnessing the interactions between light and matter has long been a central objective in condensed matter physics and the advancement of quantum technologies. One prominent approach is the manipulation of quantum system properties through classical light, a technique known as Semi-Classical Floquet Engineering (SCFE)~\cite{Grossmann1991, lindner_floquet_2011, Gomez-Leon2013, Gomez-Leon2014, Benito2014, Bello2016, Grushin2014, Diaz2019, Rudner2020, Oka2019,  GloriaRamon1997, GloriaRamon2004, Engelhardt2016, Creffield2010}, to differentiate it from its classical counterpart known as \textit{Classical Floquet Engineering}~\cite{Higashikawa2018FloquetEO}.
In both cases, a time-periodic external field imprints a discrete time-translation symmetry in the system, which in the high-frequency regime can be approximated by a static, stroboscopic description, with the driven Hamiltonian being now renormalized by the external field control parameters~\cite{GomezLeon2011,Eckardt2015}.

A more recent paradigm, termed Quantum Floquet Engineering (QFE)~\cite{quantumtoclassical, Dmytruk2022,electromagneticTB,Eckstein2022, floquetEngKennes}, involves the use of quantum rather than classical light, in the context of what has been dubbed \textit{cavity quantum materials}~\cite{cavityquantummat, Hubener2021, Bloch2022}.
In this case, the cavity field is time-independent, but in its semi-classical limit, its intrinsic free evolution allows to reproduce several features characteristic of SCFE~\cite{quantumtoclassical}. This resemblance has in fact led to the creation of the term QFE to designate the external manipulation of quantum materials with cavities.\\
In view of these similarities, cavity quantum materials can take advantage of techniques widely used in the study time-periodic systems~\cite{gomezleon2023anomalousfloquetphasesresonance}, but also provide insight beyond SCFE, when backaction and light-matter correlations between the quantum system and the external field are not negligible~\cite{perezgonzalez2023manybodyoriginanomalousfloquet}.
In addition, a practical aspect of quantum light is the mitigation of heating~\cite{Dmytruk2022, cavityquantummat}, which is inherent to SCFE and hinders the observation of Floquet physics~\cite{DAlessio2014,Bukov2016,Weidinger2017,Abanin2015,zhang2016,Iwahori2017,McIver2020}.

As stated above, SCFE can be seen as a limiting case of QFE where the light field is macroscopic~\cite{quantumtoclassical}, which implies that back-action is negligible and light-matter correlations are washed-out. However, it is conceptually very different from QFE because it is based on a discrete time-translation symmetry imprinted by an external periodic field and the system is characterized by periodic quasienergy bands~\cite{Eckardt_2015}. In contrast, QFE refers to isolated systems, but is more challenging due to its many-body nature and to the highly non-linear dependence on the photonic operators of gauge-invariant Hamiltonians~\cite{resolutionNori, ResolutionZueco, gaugeinvZueco, giDickHopfield,Eckstein2022, quantumtoclassical, Eckstein2020, Dmytruk2022, demleradf}.
In fact, previous works in the field have proposed different methods to obtain simpler, effective Hamiltonians, such as mean-field ansatzs~\cite{floquetEngKennes,Dmytruk2022} or high-frequency expansions for the matter part~\cite{Eckstein2022,Eckstein2020,quantumtoclassical}, with the preservation of gauge-invariance becoming a guiding line.

In the well-known field of SCFE, driven systems have revealed as fruitful platforms for creating and manipulating topological properties, being one of its most celebrated manifestations the presence of gapless boundary modes that are resilient to certain perturbations, in finite-size systems. This robustness encompasses many interesting physical phenomena, as well as promising applications for quantum technologies.
Then, by designing a suitable time-periodic driving, one can either tune the topology of a system~\cite{mipaper2, Gomez-Leon2013}, induce non-trivial topological behaviour in an otherwise trivial set-up \cite{AlvaroGloriaGrafeno}, or create unique phases with no static counterpart~\cite{Rudner2013, MoraisAF, AnomalousFI,gomezleon2023anomalousfloquetphasesresonance}. The classification of these driven topological phases can be systematically carried out in terms of symmetries and dimensionality, in the spirit of the well-known tenfold-way in the static case, with the associated topological invariants~\cite{Roy2017}.

In QFE, the interplay between topology and quantized electromagnetic fields has not been explored to its fullest, and in particular, the effect of the interaction on the symmetries that support the topological phase remains an open question. This issue is not only crucial to QFE, but to any other set-up in which robust edge modes couple to quantum light, either for detection or for quantum information transfer protocols, since the breaking of a certain symmetry can jeopardize topological properties arising in the interacting system and their robustness.

In this context, the present work has a twofold aim. On the one hand, we present \textit{a general framework to obtain a non-perturbative, polynomial expansion of the full, gauge-invariant Hamiltonian based on the truncation of higher-order photon-exchange processes}, which provides accurate results for arbitrary coupling strength. This method can successfully capture the physics of both, fermions and photons, in agreement with the predictions of gauge-invariant models. Importantly, it also keeps track of light-matter correlations, which are essential not only to correctly predict the properties of the many-body system, but also to carry out a faithful comparison between SCFE and QFE.\\
On the other hand, \textit{we investigate the specific role of light-matter correlations in QFE, for the case of topological systems}. As a proof of concept, we study the case of a Su-Schrieffer-Heeger (SSH) chain~\cite{SSH0,SSH1} coupled to a cavity.
We show how to produce edge states by tuning the light-matter coupling and the number of photons in the cavity and discuss how light-matter correlations affect the topological protection due to changes in the fundamental symmetries of the system, even for a largely-detuned cavity~\cite{breakdownprot}. This is especially relevant for the experimental realization of medium-sized topological systems, where corrections to the thermodynamic limit can become significant~\cite{romanroche2024cavityqedmaterialscomparison}.
Finally, we discuss how the spectroscopic analysis of the cavity can be used to detect our findings and the possibility of using the cavity to control the electron transfer between edge states.

\section{Model}

\subsection{Gauge-invariant light-matter Hamiltonian}
Consider a tight binding description of spinless fermions in a lattice, coupled to a single-mode cavity.
The quantized electric field in the cavity can be expressed in terms of the vector potential $\vec{A}=A_0 \left(d+d^{\dagger}\right)\hat{u}\left(r\right)$, where $\hat{u}$ is the unitary polarization vector\footnote{Notice that we have assumed the dipole approximation, where the vector potential does not vary within the scale of the  electronic system and its amplitude is independent of the spatial coordinate.}. In the Coulomb gauge, the coupling between the two systems can be introduced via the Peierls substitution (see Appendix A), yielding the Hamiltonian~\cite{gaugefixingTB}:
\begin{equation}
H=\Omega d^\dagger d+\sum_{l,j=1}^{N}J_{l,j}e^{i\phi_{l,j}}c_{l}^{\dagger}c_{j}\ ,\label{eq:PShamiltonian}
\end{equation}
with $\Omega$ is the cavity frequency, $N$ the number of sites in the array and $J_{l,j}$ the hopping.
The creation/annihilation operators $c^\dagger_j$/$c_j$ are fermionic, while $d^\dagger$/$d$ are bosonic. In Eq.~\eqref{eq:PShamiltonian}, the hopping gets dressed by the Peierls phase, $\phi_{l,j} = \eta_{l,j}\left(d^{\dagger}+d\right)$, with $\eta_{l,j} = e A_0 r_{l,j}$ acting as the effective coupling strength ($e$ is the particle charge and $r_{l,j}$ is the distance between the $l^{\mathrm{th}}$ and $j^{\mathrm{th}}$ site). Notice that since $r_{l,j} = - r_{j,l}$, the coupling $\eta_{j,l}$ depends on both the distance between sites and the direction of the hopping.\\

Eq.~\eqref{eq:PShamiltonian} represents a gauge invariant description of the system for arbitrary coupling, but it is also highly nonlinear in the photonic operators, which makes its use difficult for practical calculations.
A standard maneuver with Eq.~\eqref{eq:PShamiltonian}~\cite{Feynman-1951,Mahan} uses the Baker-Haussdorf-Campbell formula, in combination with the Hubbard operators, $Y^{n,n^\prime} = \vert n \rangle \langle n^\prime \vert $, with $n,n^\prime \in \mathbb{Z}\geq 0$ and $|n\rangle$ the number of photon states, to write the Peierls phase as:

\begin{equation}
e^{i\eta_{l,j}(d^\dagger + d)} = \sum_{n=0}^{\infty} g_{n,n}^{l,j} Y^{n,n} + \sum_{n\neq  m=0}^{\infty} g_{m,n}^{l,j} Y^{m,n}, \label{eq:Peierlsphase}
\end{equation}
where the diagonal elements $g^{l,j}_{n,n} = \langle n\vert e^{i\eta_{l,j}(d^\dagger + d)} \vert n\rangle$ are given by

\begin{equation}
    g^{l,j}_{n,n} = e^{-\eta^2_{l,j}/2}L_n(\eta^2_{l,j}),
    \label{eq:diagonalterms}
\end{equation}

with $L_n(\eta^2_{l,j})$ being the Laguerre polynomials, and the off-diagonal terms $g_{m,n}^{l,j} = \langle m \vert e^{i\eta_{l,j}(d^\dagger + d)} \vert n\rangle$ ($m\neq n$) are written in terms of the Hypergeometric functions $_{1}F_{1}(-m;1+n-m;\eta^2_{l,j})$ (see Appendix B)\footnote{Although the Krummer function is well-defined only for $m\geq s$, a numerical calculation of the exact matrix elements shows that they are symmetric $g^{j,i}_{n,m} = g^{j,i}_{m,n}$.}:
\begin{align}
g_{m,n}^{l,j} = & e^{-\eta_{l,j}^{2}/2} \frac{\left(i\eta_{l,j}\right)^{m-n}}{\left(m-n\right)!}  \nonumber \\
& \times \sqrt{\frac{m!}{n!}} \,_{1}F_{1}(-m;1+n-m;\eta^2_{l,j}).
\label{eq:krummer}
\end{align}
After all these manipulations Eq.~\eqref{eq:PShamiltonian} becomes:
\begin{align}
H =&  \sum_{n=0}^{\infty}  ( n\Omega   + \sum_{\langle j, l\rangle=1 }^{N} g_{n,n}^{l,j} J_{j,l}c_{j}^{\dagger}c_{l} ) Y^{n,n} \nonumber\\
&+ \sum_{n\neq m=0}^{\infty} \sum_{\langle j, l\rangle =1}^{N}g^{l,j}_{m,n} J_{j,l} c_{j}^{\dagger}c_{l} Y^{m,n}\ , \label{eq:totalexactH}
\end{align}
where the first line can be interpreted as a state-dependent cavity frequency shift and the second as photon-assisted hopping. Note that Eq.~\eqref{eq:totalexactH} is a non-perturbative polynomial expansion of the Peierls phase in terms of photon operators.
\begin{figure}[t]
	\centering
	\includegraphics{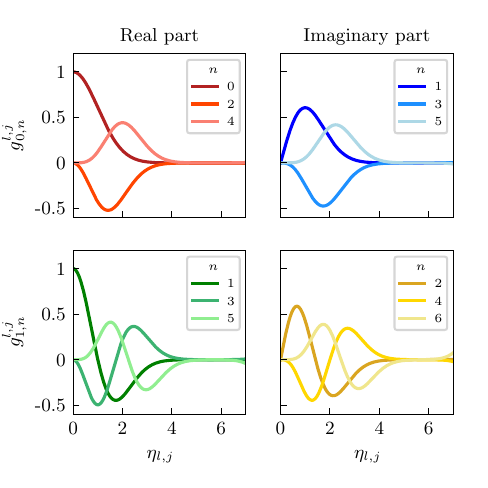}
	\caption{\label{fig:laguerrekrummer} Coefficients $g^{l,j}_{m,n}$ as a function of $\eta_{l,j}$ for different values of $m,n$. When $m = n$, the Laguerre polynomials are obtained.}
\end{figure}

It is enlightening to plot the coefficients $g^{l,j}_{n,n}$ and $g^{l,j}_{m,n}$ as a function of the coupling. While the Laguerre polynomials are purely real, the coefficients $g_{m,n}^{l,j}$ are real or imaginary depending on the values for $n$ and $m$. This is shown in Fig. \ref{fig:laguerrekrummer}, where only the non-zero contribution has been plotted for each pair $n,m$. 
Importantly, in the non-interacting limit $\eta\rightarrow 0$, the diagonal coefficients $g_{n,n}^{l,j}\rightarrow 1$ go to one, while the off-diagonal ones $g_{m,n}^{l,j}\rightarrow 0$: photon transitions disappear and we are left with the two Hamiltonians, as expected, where the energy spectrum is made of shifted copies of the fermionic energy levels, separated by $\Omega$.
The oscillating behavior of both $g_{n,n}^{l,j}$ and $g_{m,n}^{l,j}$ indicates which photonic subspace will have a larger weight depending on the coupling strength considered, with a highly non-linear dependence. Hence, the plot reveals how multi-photon processes emerge from a single-mode cavity as the coupling strength is tuned. 
Lastly, the dynamical localization prefactor $\mathrm{exp}(-\eta^2/2)$ ensures that all coefficients go to zero for large enough coupling strength, for all $n,m$.\\

The expansion in Eq. \ref{eq:totalexactH} also highlights the differences between the physical processes taking place in the total system, and that are highly concealed in the complexity of the Peierls phase: the effect of the interaction with a fixed number of photons, and that of photon-exchange processes. As we will see below, the separation of the dominant and subdominant energy scales of the system is crucial to derive  effective, simpler Hamiltonians.

\subsection{Truncation of photon exchange}

To make the Hamiltonian simpler, we propose a method to derive an effective description for the physics of the total system: to \textit{truncate the photon exchange to include only one-photon transitions}, i.e., $m = n\pm 1$ in the second term of Eq.~\eqref{eq:totalexactH}. Let us point out the following considerations for this truncation:\\

$i)$ \textbf{Regime of interest}: the truncation is specially well-suited for the high-frequency regime $\Omega \gg J^{\prime}, J$, where the hybridization between bands with a different number of photons is weak. In that case, the terms that change the number of photons in the system are small perturbations that can be treated with standard methods. This is, in fact, the regime of interest for the purposes of QFE that we will be discussing in the following. Nevertheless, our truncation scheme also works in the presence of resonances involving at most one-photon exchange. In section 3.4 we will also provide numerical evidence for this.\\ 

$ii)$ \textbf{Non-perturbative}: the truncated Hamiltonian is not equivalent to a perturbative expansion of the Peierls phase in the coupling, which would only be valid for weak interaction strengths. The non-linear nature of the Peierls phase is now encoded in the coefficients $g^{l,j}_{m,n}$ and their non-linear dependence on $\eta_{l,j}$. This ensures that the effective Hamiltonian will not break down for large values of $g/\Omega$ \cite{resolutionNori}, introducing gauge ambiguities in the description. In section 3.4 we will deepen in the comparison of our truncated Hamiltonian and the perturbative expansion of Eq. \ref{eq:PShamiltonian} to second order in $\eta$.\\

$iii)$  \textbf{Many-body physics}: as the effective, truncated Hamiltonian describes fermions coupled to photons, we can keep track of light-matter correlations. This is essential to connect with standard experimental measurements in Quantum Optics, such as the cavity transmission coefficient, where the backaction on the photonic field allows to detect properties of the electronic system.\\

\begin{figure*}
    \includegraphics{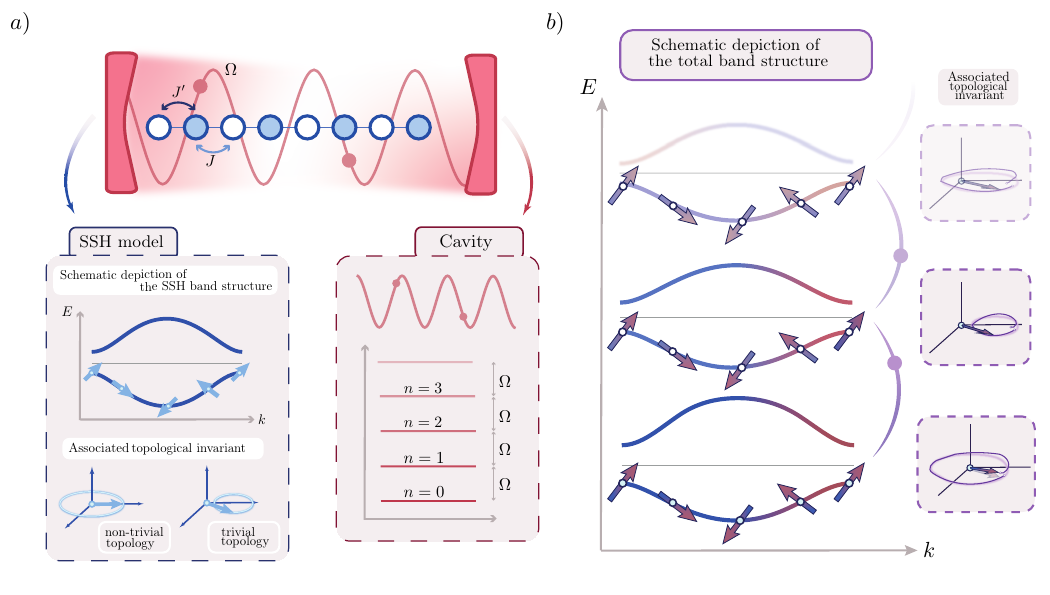}
    \caption{\label{fig:schematic} $a)$ Schematic representation of the system: an SSH chain (in blue) interacting with a photonic field in a cavity (in red). The SSH chain consists on alternating hopping amplitudes, $J^\prime$ and $J$, whose ratio define two distinct topological phases. The latter can be identified by the value of the topological invariant, which can be represented as a winding number in parameter space. The photonic field is defined by its frequency $\Omega$, and its energy levels determined by the number of photons $n$ in the cavity. $b)$ The resulting band structure of the total system corresponds to a succession of photonic subspaces where the electronic energy levels are renormalized differently. The presence of photon-exchange processes induces a coupling between Fock subspaces (purple lines), with a direct impact on the topological invariant.}
\end{figure*}

Finally, while this approach is general for arbitrary lattice systems, here we are interested in topological ones. To deepen further into the features of the truncated model, as well as the comparison between SCFE and QFE, we will particularize our results to the canonical SSH chain for the matter part (see Fig. \ref{fig:schematic}).

\section{High-frequency regime in QFE}

\subsection{Hamiltonian for the SSH chain.} 
The bipartite lattice of the SSH chain consists on an alternating pattern of different intra- and the inter-dimer hopping amplitudes, $J_{2j-1,2j} = J_{2j,2j-1} \equiv J^{\prime}$ and $J_{2j,2j+1} = J_{2j+1,2j} \equiv J $, respectively. Here, $j=1,2,...,2N$ is a site index, and $2N$ is the total number of sites, being $N$ the number of unit cells. 
This leads to the definition of two distinct sub-lattices, $A$ and $B$, and their corresponding creation/annihilation operators ($c^{(\dagger)}_{2j - 1}\rightarrow a^{(\dagger)}_p$ and $c^{(\dagger)}_{2j} \rightarrow b^{(\dagger)}_p$, where $p=1,...,N$ is a cell index). With this, the SSH Hamiltonian reads 

\begin{equation}
H_{\mathrm{SSH}} =  \sum_{p=1} \left( J^\prime a^\dagger_p b_{p} + J a^\dagger_{p+1} b_{p} + \text{h.c.} \right).
\label{eq:sshunperturbed}
\end{equation}

The SSH model is a canonical example for topological insulators in one dimension showcasing two distinct topological phases as a function of the ratio $J^\prime/J$~\cite{TopSymmetries, classTIandSC}. For the topological phase $J^\prime/J <1$, two edge states topologically protected by chiral symmetry appear within the gap~\cite{mipaper1}. The hopping dimerization can be understood as a result of the different intra/inter-dimer distance, $r^{\prime}$ and $r$ (the unit cell length is set to $r + r^\prime = 1$), and when the chain is coupled to light as in Eq. \eqref{eq:PShamiltonian}, this also gives a dimerized coupling strength that depends on either $r$ or $r^{\prime}$: $\eta^{(\prime)} = e A_{0} r^{(\prime)}$. In SCFE, the dependence on the lattice geometry has well-known consequences for the topology of the chain in SCFE, such as the possibility to produce a topological phase when the system initially is topologically trivial~\cite{Gomez-Leon2013}. For further details on the SSH Hamiltonian, the reader should go to Appendix $C$.

Chiral symmetry is important in 1D systems, because it can provide topological protection. This is the case of the SSH model, which has time-reversal, particle-hole and chiral symmetry~\cite{TopSymmetries, classTIandSC}. In particular, the later can be represented by a unitary operator $\Gamma$ that anti-commutes with the fermionic Hamiltonian. For the case of the SSH chain, the chiral operator can be expressed as $\Gamma = \sum_{p}\left(a_p^\dagger a_p - b_p^\dagger b_p\right)$ and fulfills  $\Gamma H_{\mathrm{SSH}}\Gamma = - H_\mathrm{SSH}$. This implies that the on-site potential must be uniform and that the hopping must connect sites between different sub-lattices only \cite{mipaper1}. \\

An equivalent definition for $\Gamma$ can be written in $k$ space. In this case, the bulk SSH  Hamiltonian $\mathcal{H}_{\mathrm{SSH}}(k)$ is defined as

\begin{equation}
\mathcal{H}_{\mathrm{SSH}}(k) = h(k)  a^\dagger_k b_k + h^{*}(k) b^{\dagger}_k a_k,
\label{eq:sshHamiltoniank}
\end{equation}

with $ h(k) = t^\prime e^{i k r^\prime} + t e^{-i k r}$. The eigenenergies of Eq. \eqref{eq:sshHamiltoniank} give the valence and conduction bands of the chain, namely $E_{\pm} = \pm \sqrt{J^{\prime 2} + J^2 + 2J^\prime J \cos(k0}$. The corresponding eigenstates, $\vert u_{\pm}(k)\rangle$, let us define the $\Gamma$ operator in momentum space as $\Gamma(k) = \vert u_{+}(k)\rangle \langle u_{+}(k) \vert - \vert u_{-}(k)\rangle \langle u_{-}(k) \vert$.\\

\begin{figure}
	\centering
        \includegraphics[width=1\columnwidth]{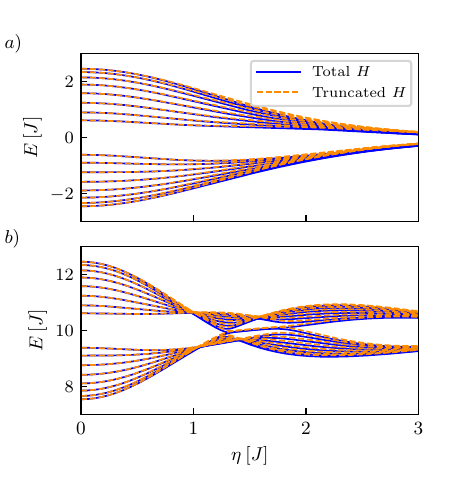}
	\caption{\label{fig:spectrums_Truncated} Energy spectrum obtained from exact, numerical diagonalization (Eq. \eqref{eq:totalexactH}) of the $a)$ zeroth, and $b)$ first photonic band, corresponding to the SSH chain in a cavity, for the total (blue) and truncated (dashed, orange) Hamiltonians. Parameters: $J^\prime = 1.5$, $J=1$, $r^\prime = 0.4$, $\Omega = 10$, $n_\mathrm{max} = 70$ (maximum number of photonic subspaces included for convergence), $N=16$. The coupling strength $\eta$ is given in units of $J$.}
\end{figure}

The presence of chiral symmetry has consequences in the energy spectrum of the system, which holds true for both the real-space and momentum-space Hamiltonians: the states come in chiral pairs, with energies $E$ and $-E$, yielding a symmetric energy spectrum. For the topological edge states, this means that they must be found at the band center ($E=0$), being degenerate in energy. However, in finite-size systems an energy splitting is expected, exponentially suppressed with the system size.

\subsection{SSH coupled to quantum light}

Fig.~\ref{fig:spectrums_Truncated} shows the spectrum of the hybrid system in the high-frequency (HF) regime $\Omega \gg J^{\prime}, J$, as a function of the light-matter coupling $\eta$ (notice that $\eta^\prime = \eta \hspace{1pt}r^\prime/r < \eta$, since $r^\prime < r$ in the trivial phase, which motivates the use of the largest coupling strength as a tuning parameter for the plots). 
Panel $a)$ ($b)$) represents the zeroth (first) photonic band. These labels refer to the fact that, for $\eta^{(\prime)} = 0$, the eigenstates of the total Hamiltonian can be factorized and the photonic wavefunction corresponds to a Fock state with a well-defined photon number, $n = 0$ ($n = 1$). However, note that when the cavity couples to the system in the off-resonant regime, the interaction will slightly modify the cavity state to a non-separable one. In both cases, the exact diagonalization of Eq.~\eqref{eq:totalexactH} and that of its truncated version, show an excellent agreement for all values of the coupling strength. Quantitative small differences between them are due to multi-photon exchange processes (see  Appendix D for a detailed analysis on the effect of higher order photon exchange).\\

The spectrum also shows the cavity-induced localization (band collapse) at large coupling, produced by the exponential suppression of the hopping in Eqs.~\eqref{eq:diagonalterms} and \eqref{eq:krummer}. 
Notice that we choose the hopping of the unperturbed SSH chain in Fig.~\ref{fig:spectrums_Truncated} such that the system is topologically trivial for $\eta = 0$. Then, while the zeroth photonic band in the $a)$ panel always lacks edge states,  the $b)$ panel shows that in the first band, increasing $\eta$ gives rise to the appearance of edge states for a certain range of values. This difference is possible in cavity quantum materials because different Fock subspaces are not identical, as it is the case of Floquet systems, where the quasienergy spectrum is made out of infinitely many replicas. The appearance of edge states as a function of $\eta$ resembles the topological phase transition of the SSH chain in the HF regime of SCFE~\cite{Gomez-Leon2013}, which takes place for certain field amplitudes. \\

\subsection{Comparison between SCFE and QFE}

The first important difference between SCFE and QFE is that the spectrum in SCFE, due to time periodicity, consists on infinite copies of identical Floquet bands. In contrast, in the present case the spectrum is bounded from below and each photonic subspace renormalizes differently, as a function of $\eta$. In fact, the photo-dressing of the electronic system under the interaction with classical and quantum light can be easily connected using the formalism of Eq.~\ref{eq:PShamiltonian}. As shown in Ref.~\cite{quantumtoclassical}, the Floquet band renormalization in SCFE can be recovered in QFE by taking the limit of large photon numbers $n\rightarrow \infty$, while keeping $\eta \sqrt{n}$ fixed. Then, the distinction between SCFE and QFE is resolved in this limit. This shows that QFE provides additional external control, as the number of photons in the cavity can now be used to manipulate the system as well. Importantly, as the spectrum of QFE is bounded from below, it has a well-defined ground state, as opposed to SCFE.

A second difference is related to the calculation of an effective matter Hamiltonian in the HF regime. Typically, in SCFE one can find a stroboscopic time-evolution operator described by an effective Hamiltonian with renormalized parameters, which can be derived by means of a Magnus expansion in inverse powers of the drive frequency~\cite{Eckardt2015}. 

In QFE, such effective Hamiltonians for either the cavity photons or the electronic system can be obtained by performing a mean field approximation, as suggested in previous works~\cite{Dmytruk2022, floquetEngKennes}. In the mean field Hamiltonians, back-action due to the interaction is contained in the renormalization of the parameters of each subsystem, while the effect of light-matter correlations is neglected. In this sense, the mean field result for QFE can be equated to SCFE, since both of them lack quantum correlations (the only difference between mean field and SCFE is the back-action from the fermions onto the cavity, due to the mean field self-consistency equations). However, we will see below why fluctuations are important and should be included in the case of QFE, especially when dealing with topological systems.

\begin{figure}[t]
	\centering
	\includegraphics{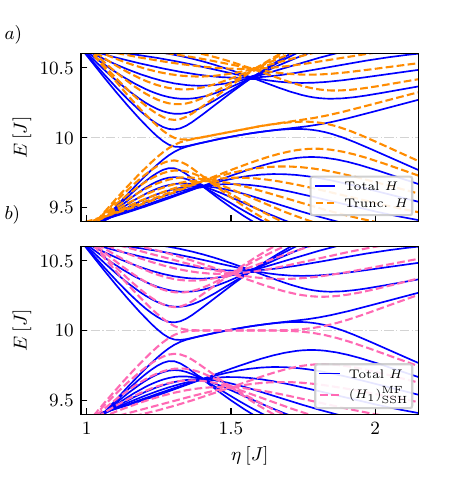}
	\caption{\label{fig:zoom_edges} $a)$ Comparison of the energy spectrum obtained for the total (blue) and truncated (orange) Hamiltonian in the first photonic band, as shown in Fig. \ref{fig:spectrums_Truncated}. We focus specifically on the energy of the edge states, and how it changes as a function of the coupling strength. $b)$ Comparison of the energy spectrum of the total Hamiltonian (blue) and the mean-field one (violet). In both plots, the dot-dashed line indicates the center of the band. The mean-field energy for the edge states remains pinned at $E = 10 [J]$, as opposed to the numerical result for the total Peierls Hamiltonian. the The parameters are chosen as in Fig. \ref{fig:spectrums_Truncated} for both subplots: $\Omega = 10$, $J^\prime = 1.5$, $J=1$, $r^\prime = 0.4$, $n_\mathrm{max} = 80$, $N =16$.}
\end{figure}

The reason is rooted in the third essential difference between SCFE and QFE, which concerns the fate of chiral symmetry in each case.\\
For SCFE in the high frequency regime, one finds that the effective Hamiltonian preserves chiral symmetry, which is why the edge states are at zero energy and topologically protected~\cite{Gomez-Leon2013}. Being the effective Hamiltonian time-independent, its symmetries can be analyzed as in any static system.\\
For the Mean Field approach to QFE, the same conclusion is obtained~\cite{Dmytruk2022}: the mean field Hamiltonian for the electronic system preserves chiral symmetry, since it retains the off-diagonal form of the unperturbed SSH Hamiltonian,

\begin{equation}
( H_n )_\mathrm{SSH}^{\mathrm{MF}} = \sum_{p=1}  \left( \tilde{J}^\prime_n a^\dagger_{p} b_{p} + \tilde{J}_n a^{\dagger}_{p+1}b_{p} + \text{h.c.} \right),
\label{eq:hMF}
\end{equation}

where $\tilde{J}^{(\prime)}_n = J^{(\prime)} \langle e^{\pm i \eta^{(\prime)} (d^\dagger + d)} \rangle$ are the renormalized hopping amplitudes (see Appendix E for further details). If one performs the MF decoupling on the truncated Hamiltonian, then the $\tilde{J}^{(\prime)}_n$ acquire the simpler form of $\tilde{J}^{(\prime)}_n = J^{(\prime)} L_n\left( \eta^{(\prime)2} \right)$. Note that there is a dependence on the photonic subspace $n$ considered. This is because an effective mean field Hamiltonian can be derived for each Fock subspace, accounting for the different dressing of the electronic band structure depending on the cavity state preparation. Importantly, the mean field Hamiltonian preserves the off-diagonal structure, which means that the chiral-symmetry operator can be defined again as $\Gamma = \sigma_z$, in the same basis as for $H_{\mathrm{SSH}}$ in Eq.~\ref{eq:sshunperturbed}.
In fact, we show the mean field result for the first photonic band in Fig.~\ref{fig:zoom_edges} (violet), as compared to the result from the total $H$ (blue) and its truncated version (orange). We focus on the region of parameters where the bandgap hosts edge states, as the breaking of the symmetry is more evident for them and has stronger consequences. In Fig.~\ref{fig:zoom_edges}, it is clear that the edge states remain pinned to the middle of the gap for the mean-field Hamiltonian, despite the increase in the coupling strength, confirming the presence of chiral symmetry and topological protection. On the contrary, for the blue and orange energy spectrum (total and truncated Hamiltonian, respectively), one can see that the pair of degenerate edge states obtained from exact numerical diagonalization, moves from the center of the gap, hybridizing with the bulk bands and indicating the breaking of chiral symmetry~\footnote{Inversion symmetry remains intact, which is why the edge states remain degenerate. This effect is analogous to that of long-range hopping in an SSH chain~\cite{mipaper1}}.

This is because QFE has an additional physical effect that is neglected in a mean field analysis and absent in SCFE: \textit{light-matter correlations}. Although they are small for $\Omega\gg J$, $J^\prime$, with topological systems one must be careful, because the topological protection of edge states might be linked to symmetries that are broken by these small corrections, as is the case of the SSH chain. The comparison with the mean-field result confirms that this is a consequence of light-matter correlations.\\

From the form of the Peierls phase in Eq.~\ref{eq:PShamiltonian}, one could think that gauge-invariant coupling to light in lattice Hamiltonians preserves chiral symmetry, because the block-structure of the matter Hamiltonian is not changed by the Peierls phase or the renormalized hopping. 
The breaking of chiral symmetry in the Coulomb gauge, and the physical process behind, must be analyzed considering both photons and fermions, i.e., taking the interacting system as a whole. Therefore, it is not enough to study the electronic part separately.\\
A complete characterization of the topological phase, as well as an analysis of the impact of light-matter correlations on topological properties, can be carried out by means of a topological invariant for the composite system. For this, let us point out that the truncated Hamiltonian captures the breaking of chiral symmetry, as shown by the numerical results, which further confirms that it can be used to study the system and its topological properties. Additionally, the speed-up in the numerical treatment of Eq.~\eqref{eq:totalexactH} compared to the exact diagonalization of the Peierls Hamiltonian in Eq.~\eqref{eq:PShamiltonian} motivates the use of the former in the following sections (we have also checked its accuracy by comparing with the full Hamiltonian).

\subsection{Comparison with the expansion of the Peierls phase and other parameter choices}

The total Hamiltonian in Eq. \ref{eq:totalexactH} is still exact and contains all orders of the vector potential, which are essential for maintaining gauge invariance for arbitrary coupling strength. Similarly, we have shown that the truncated Hamiltonian successfully reproduces the exact energy spectrum provided by the exact diagonalization of Eq.~\ref{eq:totalexactH}. This means that the truncation of photon exchange processes, under certain parameter choices, abides by the predictions of gauge-invariant models. \\ 

This consideration is crucial, since any naive approach to the problem of light-matter coupling can yield inconsistent results and gauge ambiguities, which are particularly relevant in the ultrastrong coupling regime. This is the case, for example, of trivially expanding the Peierls phase to second order \cite{manipulatingTruncation, mottpolaritons} in an attempt to mimic the structure of the light-matter Hamiltonian in the continuum, following the minimal coupling replacement $\hat{p}^2 \rightarrow \left( \vec{p} - e\vec{A} \right)^2$ ($\vec{p}$ is the particle momentum). In this expression it is easy to identify a linear term in the vector potential $\propto (d^\dagger +d)$ and a quadratic one $\propto (d^\dagger + d)^2$, the so-called paramagnetic and diamagnetic contributions. However, for the Peierls phase used in lattice models, the preservation of gauge invariance requires to keep the full dependence on the photonic operators.\\

\begin{figure}[t]
	\centering
	\includegraphics{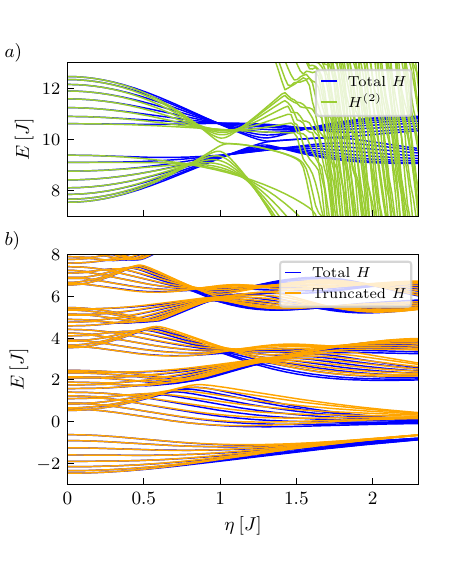}
	\caption{\label{fig:resonance_comparison} $a)$ Comparison of the energy spectrum of the total Peierls Hamiltonian and $H^{(2)}$, for $\Omega =10$, as a function of the coupling strength. $b)$ Comparison of the energy spectrum of the total Hamiltonian and the truncated one including only one-photon transitions, for a photon frequency lower than the bandwidth of the chain, $\Omega =3$. The rest of the parameters are chosen as follows: $J^\prime = 1.5$, $J=1$, $r^\prime = 0.4$, $n_\mathrm{max} = 80$, $N =16$.}
\end{figure}

The difference between the perturbative expansion of the Peierls phase and the truncated version of Eq. \ref{eq:totalexactH} lies in the coefficients $g^{l,j}_{m,n}$, which are non-linear functions of $\eta_{l,j}$, and therefore the expansion in photonic Hubbard operators is unequivalent to that of a direct series expansion in powers of $\eta_{l,j}$. Fig.~\ref{fig:resonance_comparison}~$a)$ shows the comparison between the energy spectrum for the Peierls Hamiltonian and its Taylor expasion to second order,

\begin{equation}
    H^{(2)} = \sum_{\langle l,j \rangle } \left[ J_{l,j} - \frac{\eta_{j,l }^2}{2} (d^\dagger + d)^2 \right] c^\dagger_l c_j + i(d^\dagger + d) \mathcal{J},
\end{equation}

where $\mathcal{J} = \sum_{\langle l,j \rangle} \eta_{j,l}J_{l,j} (c^\dagger_j c_l - c^\dagger_l c_j) $ contains a photon-dependent correction term to the hopping amplitudes. $H^{(2)}$ shows a good agreement for small coupling strength but differs radically from the Peierls Hamiltonian as the coupling increases. 

$H^{(2)}$ shows a good agreement for small coupling strength but differs radically from the Peierls Hamiltonian as the coupling increases. As shown in Ref. [36], the coupling strength should not be used as a perturbative parameter in the ultra-strong coupling regime, which causes the failure of the Taylor expansion and the subsequent violation of gauge invariance. This explains the mismatch with the Peierls-substituted Hamiltonian. Accuracy for larger coupling strengths would improve as higher orders of the expansion are included, yet the breakdown of the energy spectrum should still be expected at some finite value of $\eta$. Note that we have chosen a far-detuned cavity frequency, as in Fig. \ref{fig:spectrums_Truncated}.\\

Similarly, let us show the validity of the truncated Hamiltonian for a different choice of parameters, namely a cavity that is resonant, i.e., its frequency is smaller than the maximum bandwidth of the electronic system. This is depicted in Fig.~\ref{fig:resonance_comparison}~$b)$. The choice for $\Omega$  results in overlapping photonic bands, leading to a pattern of crossings and anti-crossings in the energy spectrum. Due to cavity-induced localization, the bandwidth tends to decrease as the coupling strength gets larger, which causes the initial overlap between different photonic bands to quickly disappear and drive the system out of resonance with the cavity. Hence, for very large coupling strengths, we verify that the two subsystems are effectively decoupled, even for lower cavity frequencies. The agreement between the total Hamiltonian and the truncated one is reasonably good for all coupling strengths, but specially for those values $\eta$ where only one-photon transitions participate in the resonant process. As expected, the mismatch shown in Fig.~\ref{fig:resonance_comparison}~$b)$ is larger than in Fig.~\ref{fig:spectrums_Truncated}.

\section{Topological invariant}

\subsection{Characterizing topology in terms of Green's functions}

In a non-interacting and isolated SSH chain, the topological phase is characterized
by a topological invariant $\mathcal{W}$, which is a quantized Zak phase that a state picks up when it is carried across the first Brillouin zone~\cite{Asboth}. It takes the discrete value $\mathcal{W} = 0$ in the trivial phase and $\mathcal{W} = 1$ in the topological one, and has a one-to-one correspondence with the number of pairs of edge states within the gap~\cite{bulkboundary}.

This topological invariant can be calculated as the winding number arising from the trajectory of the Bloch vector (see Fig. \ref{fig:schematic}~$a)$ for an schematic representation), but it has been shown that Green's functions (GFs) provide an alternative approach, with the possibility to incorporate many-body effects~\cite{Gurarie2011,Gurarie2011-2,Gurarie2012, Shiozaki2018, Shapourian2017}. Importantly, its value converges to the standard winding number in the non-interacting limit.
In particular, the winding number for the SSH chain in terms of Green's functions can be written as:
\begin{equation}
\mathcal{W} = \frac{1}{4\pi i} \int_{-\pi}^{\pi}dk \hspace{2pt} \Gamma [\hat{G}(k,0)]^{-1}\partial_k \hat{G}(k,0).
\label{eq:Winitial-1}
\end{equation}
where we have defined the matrix $\hat{G}(k, \omega)$ as the Fourier transform of the retarded Green function with matrix elements $G_{\alpha,\beta}(k,t)=-i\theta(t) \langle \{ \alpha_k(t), \beta_k^\dagger \} \rangle $ and $\alpha,\beta =A,\hspace{2pt}B$ the sub-lattice index. Equivalently, it can also be defined in terms of the parent Hamiltonian, which for the case of Eq. \ref{eq:Winitial-1}, is the $k$ space SSH Hamiltonian:  $\hat{G}(k, \omega) = (1 - \mathcal{H}_{\mathrm{SSH}}(k))^{-1}$. \\
Importantly, the definition of $\mathcal{W}$ from Eq. \ref{eq:Winitial-1} requires a system with chiral symmetry $\Gamma$. From these works one can also infer that for a system exhibiting chiral symmetry, such as the SSH chain, the corresponding GF will have an off-diagonal structure. This is because the GF satisfies the same symmetry constraints as the parent Hamiltonian.

\subsection{Topological invariant for the hybrid system}
Within the Green functions framework, we can extended our previous discussion and determine the effect of light-matter correlations in the topological invariant of the SSH chain.
Also, the truncated Hamiltonian will allow us to find analytical expressions for the Green's function and explicitly demonstrate the breaking of chiral symmetry due light-matter correlations.

In order to do this, we must first generalize the expression for the Green's function, to include the presence of photons from the cavity.
For that we use the fact that at high frequency, Fock subspaces are widely separated in energy.
Then, we define the fermionic Green's function, projected onto the $n$-th Fock band, as:
\begin{equation}
G^{n,n}_{\alpha,\beta}(k,t) = -i\theta(t) \langle \{ \alpha_k(t)Y^{n,n}, \beta_k^\dagger \} \rangle 
\label{eq:Gssab-1}
\end{equation}
Notice that it is now a many-body Green's function with fermionic and photonic components. Importantly, the projection to a photon subspace $Y^{n,n}$, will allow us study the gap closures and change in topology, as a function of $\eta$, in the different Fock subspaces.\\

Let us first, show how the non-interacting limit is recovered from Eq.~\ref{eq:Gssab-1}. For $\eta^\prime = 0$, the distinct Fock subspaces are decoupled, and the energy spectrum consists on infinitely many copies of the electronic band structure separated by an energy splitting of $\Omega$. Each eigenstate has a well-defined photon number $m$, and can be factorized between a fermionic part $\vert \psi(k)\rangle$ and a photonic one $\vert m \rangle$, yielding $\vert \Psi (k)\rangle = \vert \psi(k)\rangle \otimes \vert m \rangle$. Therefore, in this limit, if one wishes to calculate $G^{m,m}(k,\omega)$, only the eigenstates such that $\langle m\vert \Psi_i\rangle  = 1$ will give a non-zero contribution as if we were considering an isolated SSH chain.

A similar conclusion can be obtained for $\eta^\prime \neq 0$, in the case of the mean field Hamiltoninan, which depends on the photon index due to the cavity state, but does not include any photon-exchange terms.
The calculation of the corresponding Green's function shows that it takes a chiral form that commutes with $\Gamma = \sigma_z$ when evaluated at $\omega = 0$, as expected:
\begin{figure*}[t]
    \centering
    \includegraphics{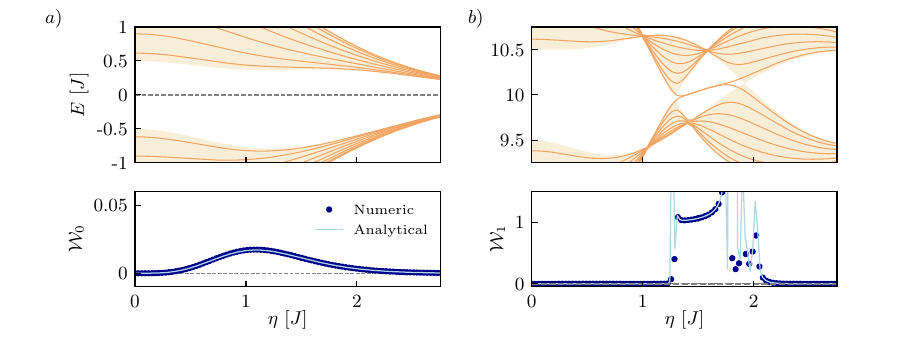}
    \caption{\label{fig:topinvariant} Loss of chiral symmetry due to electron-photon correlations, as reflected in the energy spectrum and the topological invariant of each photonic band: $a)$ $n = 0$, and $b)$ $n = 1$. The shaded regions in the top plots indicate the width of the corresponding energy band in the thermodynamic limit, while the horizontal, gray line corresponds to the center of the band. In the bottom plots, the topological invariant is shown for both the numerical and the anaylitical solution of $G^{n,n}(k,\omega)$. $J^\prime = 1.5$, $J=1$, $r^\prime = 0.4$, $n_\mathrm{max} = 80$, $N =16$.}
\end{figure*}
\begin{equation}
\hat{G}_{\mathrm{MF}}^{n,n}(k, \omega) = \frac{1}{2\pi} \frac{1}{\omega^2 - \varepsilon^2_{n}(k)} \left( 
\begin{array}{cc}
\omega & h_{n}(k) \\
h^{*}_{n}(k) & \omega 
\end{array}
\right),
\end{equation}
where $\varepsilon^2_n = h^{*}_{n}(k) h_{n}(k)$ corresponds to the photo-dressed electronic band structure, with $h_n (k) = \tilde{J}^\prime_n + \tilde{J}_n e^{ik}$. For this reason, the calculation of the winding number for each photonic band, which can also be generalized to
\begin{equation}
\mathcal{W}_n =  \frac{1}{4\pi i} \int_{-\pi}^{\pi}dk \hspace{2pt} \Gamma [G^{n,n}(k,0)]^{-1}\partial_k G^{n,n}(k,0),
\label{eq:Ws}
\end{equation}
perfectly predicts the existence of topological edge states from the mean field Hamiltonian. Its value not only depends on the coupling $\eta$, due to the dressing of the electronic band structure, but also on the band index $n$, as predicted from the mean field Hamiltonian.

Including light-matter correlations due to photon exchange processes complicates the calculation, but for the truncated Hamiltonian one can show analytically that the corresponding Green's function takes a non-chiral form, with no operator satisfying $\Gamma \hat{G}^{n,n}(k,0) \Gamma =-\hat{G}^{n,n}(k,0)$. We refer the interested reader to the Appendix F, where all details regarding the extensive calculation of $\hat{G}(k,0)$ are provided. From this result we can conclude that light-matter correlations, induced by the exchange of photons between different bands, are responsible for the breaking of chiral symmetry in the interacting system. When the electronic part has topological properties, this is of a crucial importance: \textit{it means that topological protection from chiral symmetry will be lost due to the correlations with the quantized photon field}. In consequence, the corresponding winding number losses its quantization, indicating that no longer is the correct topological invariant for the system.\\

This is illustrated in Fig. \ref{fig:topinvariant}, where we plot $\mathcal{W}_0$ and $\mathcal{W}_1$ (lower panels) together with the detail of the asymmetry induced by the coupling with the cavity photons in the energy spectrum (upper panels). We compare the full numerical calculation of the winding number, to the one obtained from the analytical expression of $\hat{G}$ in Appendix F. We omit here the full solution for clearness, since we are mainly interested in its structure and to use it as a double-check for numerics.\\
First, let us remark that the agreement between the numerical and the analytical solution is excellent. Importantly, both $\mathcal{W}_0$ and $\mathcal{W}_1$ capture the absence of chiral symmetry through the loss of quantization: note how the asymmetry in the spectrum is proportional to the deviations from integer values in the topological invariant, in both cases. This is specially evident in panel $a)$, where the maximum asymmetry corresponds to the maximum value of $\mathcal{W}_0$. Similarly, in the topological region of panel $b)$, $\mathcal{W}_1$ takes the closest values to $1$ when the deviation of the energy are minimal, i.e., chiral symmetry is weakly broken. Besides, the values of $\mathcal{W}_{0,1} \sim 0$ or $1$ are linked to the presence of edge states, and reflect the gap closures in the system.

\section{The breaking of chiral symmetry}
\subsection{Is the hybrid system topological?}
Our findings, supported by both analytical and numerical results, confirm that the original chiral symmetry of the SSH chain is broken once it couples to the cavity, for arbitrary coupling strength.
This is a consequence of light-matter correlations and it is not captured by mean field theories, as shown in Fig.~\ref{fig:zoom_edges}, where the truncated Hamiltonian reproduces the bands asymmetry (orange), while the mean field result does not (dashed, violet lines).\\
This also proves that the truncation of Eq.~\eqref{eq:totalexactH}, to only include one-photon processes, simplifies the light-matter Hamiltonian substantially, keeps numerical accuracy, and points out the strengths and limitations of the mean field treatment.
Hence, while mean field results provide a reasonably good starting point for calculations involving strong light-matter interactions, light-matter correlations can be crucial in some cases. In these situations a more complex analysis can be carried out using our truncated Hamiltonian for increased efficiency.\\

However, it is true that the MF Hamiltonian does reproduce the appearance and disappearance of edge states in each gap as a consequence of the renormalized ratio between the intra- and inter-dimer hopping. Therefore, it seems appropriate to ask: \textit{does that mean that the system can be regarded as topological?}\\
While this might seem paradoxical, it is important to note that the corrections due to light-matter correlations, breaking chiral symmetry, represent a perturbative energy scale compared to the hopping. Hence, the renormalization of the hopping amplitudes controls the appearance/disappearance of edge states as a function of the coupling. However, their topological protection by chiral symmetry is irrevocably lost due to light-matter correlations.\\

There are other examples in the literature where systems exhibit edge states but showcase no topological protection. This is the case of the extended SSH model, with long-range hopping, studied in Ref.~\cite{mipaper1}. For the case of first- and second-neighbour hopping, chiral symmetry is broken by the later, yet there are gaped configurations where the presence of absence of edge states can be directly linked with the band topology given by first-neighbour hopping amplitudes.\\
Recently, some works have demonstrated that generalized Zak phases for many-body systems can provide quantized topological invariants for the SSH chain embedded in a single mode cavity~\cite{nguyen2024electronconductancecavityembeddedtopological,perezgonzalez2023manybodyoriginanomalousfloquet}.
We must stress that here we discuss the breakdown of the original chiral symmetry in the SSH chain due to light-matter correlations, but we do not explore the existence of polariton topological phases with different symmetries. Importantly, both works also stress the importance of corrections to mean field theories, which is one of the main conclusions from our results.

\subsection{Effective model}

Previously, we have shown that the chiral symmetry of the SSH chain breaks down due to light-matter correlations, and that our truncated Hamiltonian to one-photon exchange processes can capture this effect. An additional advantage of the truncated Hamiltonian is that it allows to find analytical expressions in many situations, such as the Green's function or the winding number previously calculated. They allowed us to explicitly demonstrate the loss of quantization of the winding number and the loss of chiral form in the Green's function.\\
Another interesting application is to derive an effective Hamiltonian for a particular Fock subspace $n$, that incorporates the effect of light-matter correlations from neighboring subspaces using our truncated Hamiltonian. In this section we use the projectors method to calculate an effective Hamiltonian in the high-frequency regime, that incorporates the effect of one-photon exchange processes.
In general, the effective Hamiltonian can be expressed as~\cite{gomezleon2024highqualitypoormansmajorana}:
\begin{equation}
    H_{\text{eff}}\left(n,t\right)=PH_{0}P-i\int_{0}^{t}d\tau PH_{1}e^{-iH_{0}\tau}H_{1}e^{iH_{0}\tau}P, \label{eq:EffectiveH1}
\end{equation}
where $H_0$ corresponds to the unperturbed Hamiltonian, that we take as the diagonal part of Eq.~\eqref{eq:totalexactH} (first line with $n=m$ terms), $H_1$ to the perturbation connecting different Fock subspaces, that we restrict to one-photon exchange processes, in accordance with our truncated Hamiltonian ($n=m\pm 1$ in the second line of Eq.~\eqref{eq:totalexactH}) and $P$ is the projector onto our subspace of interest, in our case, the Fock subspace band with $n$ photons. 
Note that the effective Hamiltonian is in general time-dependent. However, it can be reduced to a time-independent expression in the high frequency regime. To show this one just needs to work in the eigenstates basis of $H_0$
\begin{align}
    H_{\text{eff}}\left(n,t\right)=& PH_{0}P \nonumber\\
    &+\sum_{\alpha,\beta}PH_{1}P_{\alpha}H_{1}P_{\beta}P\frac{e^{-i\left(E_{\alpha}-E_{\beta}\right)t}-1}{E_{\alpha}-E_{\beta}}
\end{align}
where $P_\alpha$ is the projector onto the subspace with eigenvalue $E_\alpha$. Finally, noticing that $E_\alpha-E_\beta \approx \pm\Omega$ for a perturbation connecting different Fock subspaces, the time-dependent terms will average to zero at relevant time-scales and can be neglected, finally arriving at:
\begin{equation}
    H_{\text{eff}}\left(n\right) \approx PH_{0}P-\sum_{\alpha,\beta}\frac{PH_{1}P_{\alpha}H_{1}P_{\beta}P}{E_{\alpha}-E_{\beta}} \label{eq:EffectiveH2}
\end{equation}

\begin{figure}
    \centering
    \includegraphics{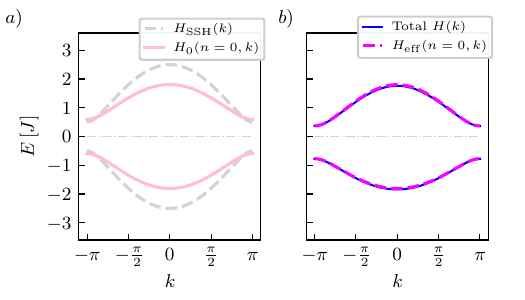}
    \caption{Comparison of the band structure of the light-coupled SSH chain, as obtained from different Hamiltonians. $a)$ Band structure obtained from: $i
    )$ $\mathcal{H}_{\mathrm{SSH}}(k)$ (dashed, gray line), the unperturbed SSH chain Hamiltonian in $k$ space, and $ii)$ $H_0(n=0,k)$, the diagonal part of Eq. \eqref{eq:totalexactH}. Both band structures are symmetric about the gap center. $b)$ Band structure obtained from: $i)$ the total Peiersl-substituted light-matter Hamiltonian $H(k)$ (blue line), and $ii)$ the effective Hamiltonian $H_\mathrm{eff}(n = 0, k)$ derived from Eq. \eqref{eq:EffectiveH1}. The effective Hamiltonian captures band asymmetry around the gap produced by light-matter correlations. All parameters are chosen as: $J^\prime = 1.5$, $J=1$, $r^\prime = 0.4$, $n_\mathrm{max} = 80$, $N =16$, $\Omega =10$, $\eta = 1$. }
    \label{fig:SpectrumEffective}
\end{figure}

In general, Eq.~\eqref{eq:EffectiveH2} can be analytically calculated and confirms that chiral symmetry for each Fock subspace is broken. In Fig.~\ref{fig:SpectrumEffective} we plot the effective band structure in momentum space, as obtained numerically from the eigenvalues of $H_{\mathrm{eff}}(n = 0,k)$. First, subplot $a)$ corresponds to the band structure of the unperturbed SSH model (dashed, grey line) and that of the diagonal part of $H$ for the Fock subspace $n=0$, $H_0(n = 0, k)$. It was obtained for $\eta = 1$, which explains the reduced maximum band width due to cavity/induced localization. Importantly, both sets of curves are symmetric about $E=0$, which indicates that chiral symmetry is preserved. This is expected, since $H_0(n = 0, k)$ lacks light-matter correlations induced by photon transitions. Similarly, subplot $b)$ shows the band structure as obtained from $H$ (with the fermionic operators transformed to $k$ space) and $H_{\mathrm{eff}}(n = 0,k)$. Note the good agreement between both and the asymmetry around the band gaps, which is produced by the lack of chiral symmetry due to one-photon transitions. \\
\subsection{Symmetry considerations in the dipole gauge}
To conclude, the breaking of chiral symmetry is also evident if one transforms Eq. \ref{eq:PShamiltonian} to the dipole gauge (SI), which yields
\begin{eqnarray}
H_{\mathrm{DG}} & = & \Omega d^\dagger d + H_{el} \\ & & + i e A_0 \Omega (d - d^\dagger )\sum_i r_i c^\dagger_i c_i \nonumber \\
& & + e A_0\Omega \left(\sum_{i} r_i c^\dagger_i c_i \right)^2. \nonumber
\label{eq:dipolegaugeS}
\end{eqnarray}
While gauge invariance ensures that the physical results are independent of the choice of gauge, the form of the light-matter coupling is gauge dependent. In the previous expression, the interaction with photons in the cavity shifts the energy of each site of the chain (second line) through the action of the displacement field $\Pi \propto i(d-d^\dagger)$. Also, an additional photon-assisted density interaction term (third line) that affects only the electronic subsystem must be included to keep gauge invariance for arbitrary coupling strength~\cite{relevanceP2}. Therefore, the presence of a photon-induced on-site potential in the fermionic system, either through the creation and destruction of photons (second line) or through a self-interaction term (third line), indicates that chiral symmetry is broken in the hybrid system. In fact, there is not an operator $\Gamma$ satisfying $\Gamma H_{\mathrm{DG}} \Gamma = - H_{\mathrm{DG}}$.

It would be reasonable to ask why one would prefer to work in the Coulomb gauge, rather than in the dipole gauge, considering the complexity of the Peierls phase. In the Coulomb gauge, the photonic vector potential enters as a phase dressing the hopping amplitudes. This is highly practical, as the coupling to light does not introduce additional matter terms. In contrast, the dipole gauge involves the electron-electron interaction term, depending on the density of charge, which complicates the evaluation of the photo-dressing of the hopping amplitudes. This aspect, which is inherent to the Coulomb gauge is one of the key principles behind controlling and manipulating matter using light, both classical and quantum. This clarity is crucial for establishing comparisons between Quantum Floquet Engineering and Semi-Classical Floquet Engineering.
\subsection{Thermodynamic limit}
The thermodynamic limit in cavity materials requires to consider the scaling of different quantities. First, the chain length and the cavity volume typically are considered as a single parameter due to the fact that an increase in the system length requires an increase in the cavity volume that contains it. Physically, the increase in the cavity volume reduces the confinement of the electric field and decreases the light-matter coupling to each particle, leading to the well-known scaling $1/\sqrt{N}$ of the coupling~\cite{Hepp1973,Wang1973}. Second, the increase number of particles leads to a collective coupling that in many cases helps to achieve strong coupling, compensating the weak single particle coupling. Then, the thermodynamic limit corresponds to taking the number of particles and the volume to infinity, while keeping the density fixed.
In this limit it has been rigorously shown that mean field Hamiltonians provide a good description for the system, indicating that fluctuations are suppressed but that the Hamiltonian is modified by the cavity due to backaction~\cite{RomanZueco2022,romanroche2024linearresponsetheorycavity,romanroche2024cavityqedmaterialscomparison,nogoTruncation}.\\

In this work we consider a single particle populating the SSH chain, which couples to a single mode cavity. Increasing the chain length involves an increase in the cavity volume and the corresponding decrease $1/\sqrt{N}$ in the light-matter coupling. Then, as for a single particle one cannot produce a collective coupling, in the thermodynamic limit the Hamiltonian effectively reduces to that of the unperturbed SSH chain, as shown in ref.~\cite{Dmytruk2022}. For this reason, and due to the fact that edge states appear at the edges of finite systems, it is interesting to study finite SSH chains at fixed light-matter coupling, where deviations from the thermodynamic limit can be important.

\section{Detection of light-matter correlations}
A well-known experimental observable in cavity QED is the frequency shift produced by the interaction with the quantum system, which can be used to perform quantum non-demolition measurements~\cite{Blais2004,Lupascu2007,Nakajima2019,GomezLeon2022}.
Its value can be extracted from the photon spectral function $A(\omega)=-\frac{1}{\pi}\text{Im} \int dt e^{-i\omega t} D(t)$, which can be experimentally measured by detection of the photons in the cavity, being $D(t)=-i\theta (t) \langle[d(t),d^{\dagger}] \rangle$ the photonic Green's function (GF).\\

\begin{figure}
	\centering
	\includegraphics{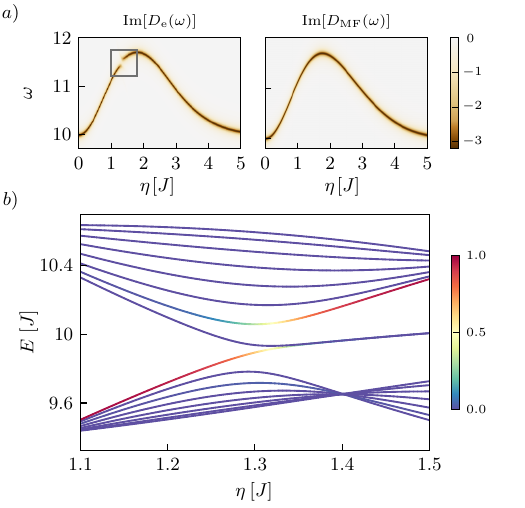}
	\caption{ \label{fig:groundstate} $a)$ Exact (left, obtained with the truncated Hamiltonian) and mean field (right) value of $A(\omega)$ as a function of $\eta$ for a cavity with $n=0$. The chain is in its ground state for a trivial phase configuration. $b)$ Truncated spectrum of the $n = 1$ subspace vs $\eta$. The color code indicates the effective cavity-mediated interaction strength $\mathcal{P}_i$ between the ground state $\vert \Psi_{\mathrm{0}}\rangle $ and each of the eigenstates $i$ in the $n=1$ subspace. Parameters: $J^\prime = 1.5$, $J=1$, $r^\prime = 0.4$, $\Omega = 10$, $n_\mathrm{max} = 70$, $N=16$. }
	\label{fig:SpectralFunction1}
\end{figure}

Now we describe how the spectroscopy of photons can be used to detect light-matter correlations and how they encode information about the edge states in the system, by analysing the behaviour of $D(\omega)$ under different parameter choices and input states.\\
In Fig.~\ref{fig:SpectralFunction1}~$a)$ we consider the cavity in its vacuum state and, to isolate the role of light-matter correlations, we compare the result obtained from the photonic GF $D_\mathrm{e}$, obtained from the truncated Hamiltonian (left) with the mean field case $D_\mathrm{MF}$  (right). It shows that, for the chain in its ground state, mean field correctly captures the frequency shift (details of the calculation in the SI):

\begin{equation}
\delta\Omega=\sum_{j,l} J_{j,l} e^{-\eta^2_{l,j}/2} \langle c_j^{\dagger} c_l \rangle,\label{eq:FrequencyShiftMF}
\end{equation}

but fails to capture the fine spectral details near $\eta\sim 1.3$, originated by light-matter correlations (framed region in the left plot).
The mechanism enhancing light-matter correlations in this region is illustrated in Fig.~\ref{fig:SpectralFunction1} $b)$, which shows the truncated spectrum in the first photonic band, as the light-matter coupling increases. The color code indicates the effective cavity-mediated interaction between the ground state of the system, $\vert \Psi_0\rangle$, and each of the eigenstates $i$ in the first photonic band ($\langle n \rangle \approx 1$), $\vert \Psi^{(1)}_i \rangle$ $(i=1,...,2N)$, calculated as $\mathcal{P}_i = |\langle \Psi_0 \vert d \vert \Psi^{(1)}_i \rangle |^2$.\\

Initially, the interaction with the first photonic band is mainly with the state on top of the valence band. However, as $\eta$ increases, this state becomes an edge state near $\eta\sim 1.3$ and since the edge states are exponentially decoupled from the bulk~\cite{TopologyDet}, the interaction jumps to the state at the bottom of the conduction band, producing the correction to the mean field result (see Fig.~\ref{fig:SpectralFunction1}).\\

\begin{figure}
	\centering
    \includegraphics{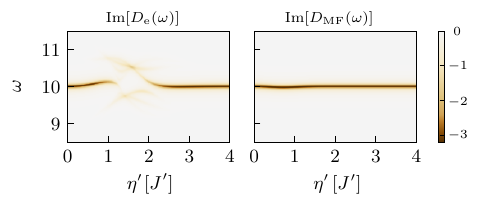}
	\caption{$A(\omega)$ vs $\eta^{\prime}$ for an empty cavity, setting $n=0$. The left plot shows the exact spectral function, while the right plot shows the mean field solution. The chain is prepared in an edge state. Parameters: $r^\prime = 0.6$, $\Omega = 10$, $n_\mathrm{max} = 70$, $N=16$, $J^\prime = 1$, $J=1.5$.}
	\label{fig:SpectralFunction2}
\end{figure}

Importantly, if instead the chain is in the topological phase for $\eta=0$, the relation between the topological phase transition and the enhancement of light-matter correlations in $A(\omega)$ still holds. Thus, it can be verified regardless of the chain being prepared in either its trivial or topological phase. Let us now consider the case of the chain being initially prepared in one of its edge states, in the topological phase. The result for $A(\omega)$ is shown in Fig.~\ref{fig:SpectralFunction2}, considering the subspace of $n=0$ photons (note that for the topological phase, the edge states in this photonic band do not disappear as a function of $\eta$, as in Fig. \ref{fig:spectrums_Truncated}).

In this case, as the edge states are exponentially decoupled from the bulk, the cavity frequency does not shift~\cite{TopologyDet}. This is very clearly shown in the right subplot, for the MF approach. However, the upper photonic band is undergoing a phase transition as $\eta^\prime$ changes: the edge states disappear for $\eta^\prime \approx 1$ and appear again for $\eta^\prime \approx 2$, showcasing the opposite behaviour of Fig. \ref{fig:spectrums_Truncated}$b)$. This is in fact reflected in the spectral photonic function. When the edge states in the first photonic band merge with the bulk, correlations strongly affect the mean field value of $A(\omega)$, until the band becomes topological again. This means that we can use the cavity, not only to externally tune the presence of edge states but also to detect the presence of edge states in subspaces with a different number of photons using spectroscopy.\\

\begin{figure}
    \centering
    \includegraphics[width = 0.95\columnwidth]{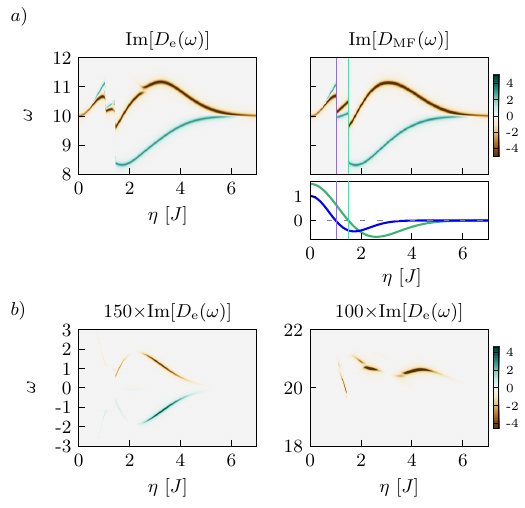}
    \caption{\label{fig:Gp1}Photonic spectral function $A(\omega)$, as obtained from $\mathrm{Im}\left[ D_{\mathrm{MF}}\right]$ and $\mathrm{Im}\left[D_{\mathrm{e}}\right]$ for $n=1$ as a function of $\eta$ and $\omega$. The fermionic system is prepared in the ground state for a trivial chain. Panel $a)$ shows the renormalization of the cavity frequency and the appearance of additional excitations, while the bottom plot shows the connection between the abrupt changes in the poles and the hopping renormalization. Panel $b)$ shows additional light-matter resonances that appear in $D_{\mathrm{e}}$ and that are absent in the MF solution, $J^\prime = 1.5$, $J=1$, $r^\prime = 0.4$, $\Omega = 10$, $n_\mathrm{max} = 70$, $N=16$.}
\end{figure}

Let us now consider a different state preparation with a non-empty cavity, with $n = 1$ photons, which is depicted in Fig. \ref{fig:Gp1}. Panel $a)$ shows the comparison between the MF and the exact result, $D_{\mathrm{MF}}(\omega)$ and $D_{\mathrm{e}}(\omega)$, respectively. While the photon frequency is shifted as a function of the coupling as in the empty case, there are extra effects that are missing in Fig. \ref{fig:SpectralFunction1} ($n=0$): the appearance of an additional pole for $\eta > 0$. This behaviour is well-reproduced by $D_{\mathrm{MF}}(\omega)$, which indicates that the two poles correspond to the exchange of virtual photons with the upper $(n+1)$ and the lower photonic band $(n-1)$, and are not caused by electron-photon correlations (this is also consistent with the fact that for the zeroth photonic band there is only one pole in the spectral function). Since each band renormalizes differently, the exchange of photons with the upper and lower band are not equivalent processes, and therefore the corresponding poles have their own dependence on $\omega$ and $\eta^\prime$. For large $\eta^\prime$, they both suppress to zero, as for the empty cavity. \\

There is an additional feature in Fig. \ref{fig:Gp1} that is also missing Fig. $\ref{fig:SpectralFunction1}$, and that appears in both the MF and the exact calculation: the presence of abrupt changes in the poles, which are related to the hopping renormalization as a function of $\eta^\prime$. This can be easily connected with the dressing of the hopping amplitudes and how they change as a function of $\eta^\prime$ (bottom plot in $a)$ panel). The oscillation of the Laguerre polynomials with $\eta^\prime$ for the non-empty cavity implies that the hopping amplitudes will be effectively suppressed at certain values of the coupling strength, with the subsequent change of sign. While the topology of the chain is determined by the absolute value of the ratio between $t^\prime g^{\prime}_{nn}$ and $t g_{nn}$, these changes in the hopping amplitudes are also reflected in the spectral function.\\

Lastly, panel $b)$ shows additional light-matter resonances that appear in the exact spectral function and that are absent in the MF solution. 

\section{State transfer dynamics}
As we discussed above, in the absence of chiral symmetry, the edge states in the SSH chain loose their topological protection, but this does not mean that they cannot be used for practical applications. 
It has been proposed that, as they are exponentially localized states, their overlap between the two ends of the chain can be used in quantum state transfer~\cite{Leijnse2011,Lang2017,Bello2016,Miguel2017}.\\

Now we show that in QFE, the cavity can be used to control charge dynamics by creating edge states in a trivial chain if $n>0$. This allows to implement state transfer protocols in systems that originally lack edge states, and fine-tune the transfer time between the two ends of the chain.\\
\begin{figure}
	\centering
        \includegraphics{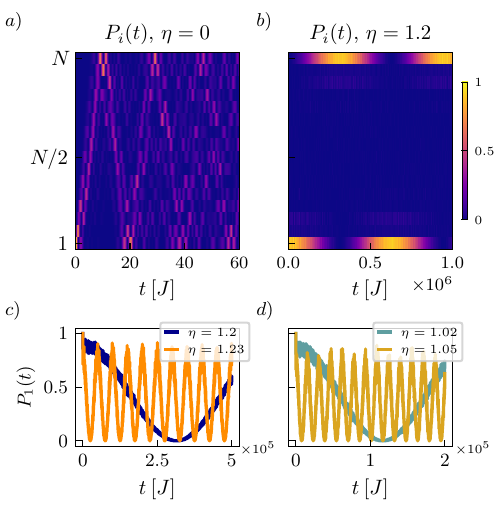}
	\caption{$a)$ and $b)$, fermionic occupation probability $P_i(t)$ of each site of the chain $i=1,...,N$ as a function of time, for a particle initially occupying the first site, $n=2$, with 
 $\eta=0$ and $\eta = 1.2$. For $\eta = 0$, the chain lacks edge states. For $\eta = 1.2$, the light-induced edge states are responsible for the oscillation of the charge occupation between ending sites. $c)$ $P_1(t)$ (on the first site of the chain) as a function of time, for $n=2$ and two different values of the coupling strength, $\eta=1.2$ and $1.23$. $d)$ $P_1(t)$ as a function of time, for $n=3$ and two different values of the coupling strength, $\eta=1.02$ and $\eta=1.23$. The rest of the parameters are chosen as: $J^\prime = 1.5$, $J = 1$, $r^\prime = 0.4$, $N = 16$, $\Omega = 10$, $n_{\mathrm{max}} = 50$.}
	\label{fig:DynamicsDispersive}
\end{figure}
This is depicted in Fig.~\ref{fig:DynamicsDispersive}~$a)$, where we plot the time evolution of the occupation probability of each site of the chain, $P_i(t)$, for a particle initially occupying the first site. If the chain is in the trivial topological phase, the particle will populate all sites of the chain, spreading along the bulk. In contrast, panel $b)$ shows that if the chain is coupled to a cavity with $n=2$ photons, there are coherent oscillations between the two ends, confirming the presence of cavity-induced edge states. (Note the change in the time scale between $a)$ and $b)$ panels). This dynamics indicates that, although chiral symmetry is broken, the transfer between the two ends of the chain still is efficient.\\

In the hybrid system, we can manipulate the transfer dynamics in two ways through the cavity: first, by modifying the light-matter coupling, and second, by changing the number of photons. This is what we show in Fig. \ref{fig:DynamicsDispersive}, $c)$ and $d)$ panels. In $c)$, a slight change in $\eta$ reduces the frequency of the oscillation, while the initial number of photons $n=2$ is fixed. Similarly, in $d)$, we change the cavity state preparation and set $n=3$ photons, for which we can also find values of $\eta$ that lead to the appearance of edge states (although, again, the unperturbed hopping amplitudes are chosen such that the isolated chain is trivial). Not only the number of photons affects the band structure and, therefore, the splitting of the edge states, but we can also simultaneously tune the coupling strength to find desired transfer periods. \\

\section{Conclusions}

We have shown that Quantum Floquet Engineering (QFE) in cavity materials is more complex than its classical counterpart, even in the case of large cavity frequency, where one would naively expect that the cavity photons can be easily traced-out and light-matter correlations neglected, without relevant consequences. 

Our work deals with the general question: what is the role
of light-matter correlations in Quantum Floquet Engineering, and how does it allow us to go beyond
Semi-Classical Floquet engineering (SCFE)? This is in fact a relevant question, since light-matter correlations are, by
definition, absent in the classical case, and can therefore set apart both cases.
While some connections between the SCFE and QFE have been previously explored \cite{Dmytruk2022,quantumtoclassical}, the particular role of light-matter correlations has been so far neglected in this context. For this, we have developed a non-perturbative truncation method to reduce the highly non-linear gauge-invariant Hamiltonian, to a simpler one that remains valid for arbitrary coupling and correctly detects changes in the symmetries that are relevant for the topology.

While this truncation scheme is general and valid for arbitrary lattice systems, we have considered a SSH chain coupled to the cavity to investigate the interplay between light-matter interaction and topological properties~\cite{PhysRevB.84.195413, PhysRevB.99.235156, breakdownprot, Dmytruk2022}. We have shown that the interaction with the cavity can control the existence of edge states via the coupling strength and the number of photons in the cavity. The later is a tuning parameter absent in SCFE.

Importantly, using the SSH model as a benchmark,  we demonstrate that light-matter interactions, quantum correlations and symmetries showcase an interesting interplay in these hybrid systems, even in the high frequency regime (this is in contrast with the high frequency regime of SCFE~\cite{Gomez-Leon2013}, where the coupling to light conserves the symmetries of the SSH chain).
The crucial point in our analysis is that photon-assisted hopping introduces quantum light-matter correlations that are not captured by a mean-field approach and break the original chiral symmetry. This is reflected in the band asymmetry and the shift in the edge states energy that the mean-field result does not capture. Therefore, it is not the hopping renormalization that shows the symmetry-breaking mechanism, but rather the presence of quantum many-body effects. 
The identification of a symmetry-breaking mechanism due to quantum fluctuations in the system could set an important milestone in the field of cavity- quantum materials and can have important implications for the detection and generation of topological phases in hybrid
light-matter platforms. In fact, this is not only important for topological insulators, but also for topological superconducting systems. In this later case, the coupling with light has also been explored as an alternative route to elucidate the existence of Majorana states
which could be jeopardized by the presence of symmetry-breaking processes~\cite{PhysRevLett.118.126803, PhysRevB.107.115418, PhysRevLett.109.257002}.

In addition, we have found that light-matter correlations can be experimentally detected in the photon spectral function, and that they contain information about the presence of cavity-induced edge states in the system. This is a consequence of the different response of bulk and edge states to the presence of cavity photons~\cite{TopologyDet}. Finally, we have shown that the cavity also provides us with a way to control the state transfer between edges of the chain. It can be used to create or destroy the edge states, as well as to tune the transfer time between the two sides.\\

Finally, we would also like to emphasize that the coupling strengths currently achieved in experimental setups are sufficiently large to observe the effects we predict in our work. Superconducting circuits stand out as promising candidates for the implementation of scalable qubit lattices and hybrid devices in combination with microwave resonators. In particular, there are many examples of ultrastrongly-coupled superconducting qubits to quantum light \cite{ultrastronglySCQ, uscscq, uscscq2}, which easily cover the range of coupling strength values at which the most notable effects occur. Notably, there are also experimental implementations of both the 1D SSH chain \cite{expsshsq, sshLCcircuit} and the 2D SSH lattice \cite{2Dsshscq} based on superconducting circuits, which could be integrated in a microwave resonator to test our results.\\
This issue could be also explored with ultracold atoms embedded in high-finesse resonators \cite{Maschler2008}. In this context, there have been several implementations of the SSH chain \cite{sshcoldatoms, Meier2016, sshRydberg}, including the measurement of the associated topological invariant. One could think of a possible set-up with atoms pinned by an external confinement potential, interacting with light inside the cavity \cite{coldatomlatticecavityqed, topinsincoldatomcavity, boseglassphasescavityqed, manybodyquantumlight, coldatomlatticecav, longrangequasipqed, coldatomcavitypot, higherorderPeierls, cavitybackaction}, which also have been implemented successfully \cite{Landig2016, mottinsuldickehbb}. \\

As an outlook, we can envision future research lines derived from our findings. For example, the use of the formalism to study systems with electron-electron interaction, or the role of dissipation.
	
\textbf{Acknowledgments}\\
G.P. and B.P.G. acknowledge the Spanish Ministry of Economy and Competitiveness for financial support through the grant: PID2020-117787GBI00. A.G.L acknowledges support from the European Union’s Horizon 2020 research and innovation program under Grant Agreement No. 899354 (SuperQuLAN). All the authors also acknowledge support from CSIC Interdisciplinary Thematic Platform on Quantum Technologies (PTI-QTEP+).\\

\textbf{Author Contributions}\\
B. P. G. did the analytical and numerical analysis under the supervision of A. G. L. and G. P. All authors discussed and analyzed the results, and contributed to the writting of the final paper. 

\textbf{Additional Information}\\
Competing financial interests: The authors declare no competing financial interests.
\bibliographystyle{quantum}
\bibliography{peierlsbib}

\begin{thebibliography}{100}

\bibitem{Grossmann1991}
F.~Grossmann, T.~Dittrich, P.~Jung, and P.~H\"anggi.
\newblock ``Coherent destruction of tunneling''.
\newblock \href{https://dx.doi.org/10.1103/PhysRevLett.67.516}{Phys. Rev. Lett.
  {\bf 67}, 516--519}~(1991).

\bibitem{lindner_floquet_2011}
Netanel~H. Lindner, Gil Refael, and Victor Galitski.
\newblock ``Floquet topological insulator in semiconductor quantum wells''.
\newblock \href{https://dx.doi.org/10.1038/nphys1926}{Nature Physics {\bf 7},
  490--495}~(2011).

\bibitem{Gomez-Leon2013}
A.~G\'omez-Le\'on and G.~Platero.
\newblock ``Floquet-bloch theory and topology in periodically driven
  lattices''.
\newblock \href{https://dx.doi.org/10.1103/PhysRevLett.110.200403}{Phys. Rev.
  Lett. {\bf 110}, 200403}~(2013).

\bibitem{Gomez-Leon2014}
\'Alvaro G\'omez-Le\'on, Pierre Delplace, and Gloria Platero.
\newblock ``Engineering anomalous quantum hall plateaus and antichiral states
  with ac fields''.
\newblock \href{https://dx.doi.org/10.1103/PhysRevB.89.205408}{Phys. Rev. B
  {\bf 89}, 205408}~(2014).

\bibitem{Benito2014}
M.~Benito, A.~G\'omez-Le\'on, V.~M. Bastidas, T.~Brandes, and G.~Platero.
\newblock ``Floquet engineering of long-range $p$-wave superconductivity''.
\newblock \href{https://dx.doi.org/10.1103/PhysRevB.90.205127}{Phys. Rev. B
  {\bf 90}, 205127}~(2014).

\bibitem{Bello2016}
M.~Bello, C.~E. Creffield, and G.~Platero.
\newblock ``Long-range doublon transfer in a dimer chain induced by topology
  and ac fields''.
\newblock \href{https://dx.doi.org/10.1038/srep22562}{Scientific Reports {\bf
  6}, 22562}~(2016).

\bibitem{Grushin2014}
Adolfo~G. Grushin, \'Alvaro G\'omez-Le\'on, and Titus Neupert.
\newblock ``Floquet fractional chern insulators''.
\newblock \href{https://dx.doi.org/10.1103/PhysRevLett.112.156801}{Phys. Rev.
  Lett. {\bf 112}, 156801}~(2014).

\bibitem{Diaz2019}
A.~D\'{\i}az-Fern\'andez, E.~D\'{\i}az, A.~G\'omez-Le\'on, G.~Platero, and
  F.~Dom\'{\i}nguez-Adame.
\newblock ``Floquet engineering of dirac cones on the surface of a topological
  insulator''.
\newblock \href{https://dx.doi.org/10.1103/PhysRevB.100.075412}{Phys. Rev. B
  {\bf 100}, 075412}~(2019).

\bibitem{Rudner2020}
Mark~S. Rudner and Netanel~H. Lindner.
\newblock ``Band structure engineering and non-equilibrium dynamics in floquet
  topological insulators''.
\newblock \href{https://dx.doi.org/10.1038/s42254-020-0170-z}{Nature Reviews
  Physics {\bf 2}, 229--244}~(2020).

\bibitem{Oka2019}
Takashi Oka and Sota Kitamura.
\newblock ``Floquet engineering of quantum materials''.
\newblock
  \href{https://dx.doi.org/10.1146/annurev-conmatphys-031218-013423}{Annual
  Review of Condensed Matter Physics {\bf 10}, 387--408}~(2019).

\bibitem{GloriaRamon1997}
Ram\'on Aguado and Gloria Platero.
\newblock ``Dynamical localization and absolute negative conductance in an
  ac-driven double quantum well''.
\newblock \href{https://dx.doi.org/10.1103/PhysRevB.55.12860}{Phys. Rev. B {\bf
  55}, 12860--12863}~(1997).

\bibitem{GloriaRamon2004}
Gloria Platero and Ramon Aguado.
\newblock ``Photon-assisted transport in semiconductor nanostructures''.
\newblock \href{https://dx.doi.org/10.1016/j.physrep.2004.01.004}{Physics
  Reports {\bf 395}, 1--157}~(2004).

\bibitem{Engelhardt2016}
G.~Engelhardt, M.~Benito, G.~Platero, and T.~Brandes.
\newblock ``Topological instabilities in ac-driven bosonic systems''.
\newblock \href{https://dx.doi.org/10.1103/PhysRevLett.117.045302}{Phys. Rev.
  Lett. {\bf 117}, 045302}~(2016).

\bibitem{Creffield2010}
C.~E. Creffield and G.~Platero.
\newblock ``Coherent control of interacting particles using dynamical and
  aharonov-bohm phases''.
\newblock \href{https://dx.doi.org/10.1103/PhysRevLett.105.086804}{Phys. Rev.
  Lett. {\bf 105}, 086804}~(2010).

\bibitem{Higashikawa2018FloquetEO}
Sho Higashikawa, Hiroyuki Fujita, and Masahiro Sato.
\newblock ``Floquet engineering of classical systems''~(2018).
\newblock  url:~\url{https://arxiv.org/abs/1810.01103}.

\bibitem{GomezLeon2011}
\'Alvaro G\'omez-Le\'on and Gloria Platero.
\newblock ``Charge localization and dynamical spin locking in double quantum
  dots driven by ac magnetic fields''.
\newblock \href{https://dx.doi.org/10.1103/PhysRevB.84.121310}{Phys. Rev. B
  {\bf 84}, 121310}~(2011).

\bibitem{Eckardt2015}
A.~Eckardt and E.~Anisimovas.
\newblock ``High-frequency approximation for periodically driven quantum
  systems from a floquet-space perspective''.
\newblock \href{https://dx.doi.org/10.1088/1367-2630/17/9/093039}{New Journal
  of Physics {\bf 17}, 093039}~(2015).

\bibitem{quantumtoclassical}
Michael~A. Sentef, Jiajun Li, Fabian K\"unzel, and Martin Eckstein.
\newblock ``Quantum to classical crossover of floquet engineering in correlated
  quantum systems''.
\newblock \href{https://dx.doi.org/10.1103/PhysRevResearch.2.033033}{Phys. Rev.
  Research {\bf 2}, 033033}~(2020).

\bibitem{Dmytruk2022}
Olesia Dmytruk and Marco Schiro.
\newblock ``Controlling topological phases of matter with quantum light''.
\newblock \href{https://dx.doi.org/10.1038/s42005-022-01049-0}{Communications
  Physics {\bf 5}, 271}~(2022).

\bibitem{electromagneticTB}
Jiajun Li, Denis Golez, Giacomo Mazza, Andrew~J. Millis, Antoine Georges, and
  Martin Eckstein.
\newblock ``Electromagnetic coupling in tight-binding models for strongly
  correlated light and matter''.
\newblock \href{https://dx.doi.org/10.1103/PhysRevB.101.205140}{Phys. Rev. B
  {\bf 101}, 205140}~(2020).

\bibitem{Eckstein2022}
Jiajun Li, Lukas Schamri\ss{}, and Martin Eckstein.
\newblock ``Effective theory of lattice electrons strongly coupled to quantum
  electromagnetic fields''.
\newblock \href{https://dx.doi.org/10.1103/PhysRevB.105.165121}{Phys. Rev. B
  {\bf 105}, 165121}~(2022).

\bibitem{floquetEngKennes}
Christian~J. Eckhardt, Giacomo Passetti, Moustafa Othman, Christoph Karrasch,
  Fabio Cavaliere, Michael~A. Sentef, and Dante~M. Kennes.
\newblock ``Quantum floquet engineering with an exactly solvable tight-binding
  chain in a cavity''.
\newblock \href{https://dx.doi.org/10.1038/s42005-022-00880-9}{Communications
  Physics {\bf 5}, 122}~(2022).

\bibitem{cavityquantummat}
F.~Schlawin, D.~M. Kennes, and M.~A. Sentef.
\newblock ``{Cavity quantum materials}''.
\newblock \href{https://dx.doi.org/10.1063/5.0083825}{Applied Physics Reviews
  {\bf 9}, 011312}~(2022).

\bibitem{Hubener2021}
Hannes H{\"u}bener, Umberto De~Giovannini, Christian Sch{\"a}fer, Johan
  Andberger, Michael Ruggenthaler, Jerome Faist, and Angel Rubio.
\newblock ``Engineering quantum materials with chiral optical cavities''.
\newblock \href{https://dx.doi.org/10.1038/s41563-020-00801-7}{Nature Materials
  {\bf 20}, 438--442}~(2021).

\bibitem{Bloch2022}
Jacqueline Bloch, Andrea Cavalleri, Victor Galitski, Mohammad Hafezi, and Angel
  Rubio.
\newblock ``Strongly correlated electron--photon systems''.
\newblock \href{https://dx.doi.org/10.1038/s41586-022-04726-w}{Nature {\bf
  606}, 41--48}~(2022).

\bibitem{gomezleon2023anomalousfloquetphasesresonance}
{\'{A}}lvaro G{\'{o}}mez-Le{\'{o}}n.
\newblock ``Anomalous {F}loquet {P}hases. {A} resonance phenomena''.
\newblock \href{https://dx.doi.org/10.22331/q-2024-11-13-1522}{{Quantum} {\bf
  8}, 1522}~(2024).

\bibitem{perezgonzalez2023manybodyoriginanomalousfloquet}
Beatriz P{\'e}rez-Gonz{\'a}lez, Gloria Platero, and {\'A}lvaro
  G{\'o}mez-Le{\'o}n.
\newblock ``Quantum origin of anomalous floquet phases in cavity-qed
  materials''.
\newblock \href{https://dx.doi.org/10.1038/s42005-024-01908-y}{Communications
  Physics {\bf 7}, 419}~(2024).

\bibitem{DAlessio2014}
Luca D'Alessio and Marcos Rigol.
\newblock ``Long-time behavior of isolated periodically driven interacting
  lattice systems''.
\newblock \href{https://dx.doi.org/10.1103/PhysRevX.4.041048}{Phys. Rev. X {\bf
  4}, 041048}~(2014).

\bibitem{Bukov2016}
Marin Bukov, Markus Heyl, David~A. Huse, and Anatoli Polkovnikov.
\newblock ``Heating and many-body resonances in a periodically driven two-band
  system''.
\newblock \href{https://dx.doi.org/10.1103/PhysRevB.93.155132}{Phys. Rev. B
  {\bf 93}, 155132}~(2016).

\bibitem{Weidinger2017}
Simon~A. Weidinger and Michael Knap.
\newblock ``Floquet prethermalization and regimes of heating in a periodically
  driven, interacting quantum system''.
\newblock \href{https://dx.doi.org/10.1038/srep45382}{Scientific Reports {\bf
  7}, 45382}~(2017).

\bibitem{Abanin2015}
Pedro Ponte, Z.~Papi\ifmmode~\acute{c}\else \'{c}\fi{}, F.~Huveneers, and
  Dmitry~A. Abanin.
\newblock ``Many-body localization in periodically driven systems''.
\newblock \href{https://dx.doi.org/10.1103/PhysRevLett.114.140401}{Phys. Rev.
  Lett. {\bf 114}, 140401}~(2015).

\bibitem{zhang2016}
Liangsheng Zhang, Vedika Khemani, and David~A. Huse.
\newblock ``A floquet model for the many-body localization transition''.
\newblock \href{https://dx.doi.org/10.1103/PhysRevB.94.224202}{Phys. Rev. B
  {\bf 94}, 224202}~(2016).

\bibitem{Iwahori2017}
Koudai Iwahori and Norio Kawakami.
\newblock ``Stabilization of prethermal floquet steady states in a periodically
  driven dissipative bose-hubbard model''.
\newblock \href{https://dx.doi.org/10.1103/PhysRevA.95.043621}{Phys. Rev. A
  {\bf 95}, 043621}~(2017).

\bibitem{McIver2020}
J.~W. McIver, B.~Schulte, F.-U. Stein, T.~Matsuyama, G.~Jotzu, G.~Meier, and
  A.~Cavalleri.
\newblock ``Light-induced anomalous hall effect in graphene''.
\newblock \href{https://dx.doi.org/10.1038/s41567-019-0698-y}{Nature Physics
  {\bf 16}, 38--41}~(2020).

\bibitem{Eckardt_2015}
André Eckardt and Egidijus Anisimovas.
\newblock ``High-frequency approximation for periodically driven quantum
  systems from a floquet-space perspective''.
\newblock \href{https://dx.doi.org/10.1088/1367-2630/17/9/093039}{New Journal
  of Physics {\bf 17}, 093039}~(2015).

\bibitem{resolutionNori}
Omar Di~Stefano, Alessio Settineri, Vincenzo Macr{\`i}, Luigi Garziano, Roberto
  Stassi, Salvatore Savasta, and Franco Nori.
\newblock ``Resolution of gauge ambiguities in ultrastrong-coupling cavity
  quantum electrodynamics''.
\newblock \href{https://dx.doi.org/10.1038/s41567-019-0534-4}{Nature Physics
  {\bf 15}, 803--808}~(2019).

\bibitem{ResolutionZueco}
Alessio Settineri, Omar Di~Stefano, David Zueco, Stephen Hughes, Salvatore
  Savasta, and Franco Nori.
\newblock ``Gauge freedom, quantum measurements, and time-dependent
  interactions in cavity qed''.
\newblock \href{https://dx.doi.org/10.1103/PhysRevResearch.3.023079}{Phys. Rev.
  Research {\bf 3}, 023079}~(2021).

\bibitem{gaugeinvZueco}
Salvatore Savasta, Omar Di~Stefano, Alessio Settineri, David Zueco, Stephen
  Hughes, and Franco Nori.
\newblock ``Gauge principle and gauge invariance in two-level systems''.
\newblock \href{https://dx.doi.org/10.1103/PhysRevA.103.053703}{Phys. Rev. A
  {\bf 103}, 053703}~(2021).

\bibitem{giDickHopfield}
Luigi Garziano, Alessio Settineri, Omar Di~Stefano, Salvatore Savasta, and
  Franco Nori.
\newblock ``Gauge invariance of the dicke and hopfield models''.
\newblock \href{https://dx.doi.org/10.1103/PhysRevA.102.023718}{Phys. Rev. A
  {\bf 102}, 023718}~(2020).

\bibitem{Eckstein2020}
Jiajun Li and Martin Eckstein.
\newblock ``Manipulating intertwined orders in solids with quantum light''.
\newblock \href{https://dx.doi.org/10.1103/PhysRevLett.125.217402}{Phys. Rev.
  Lett. {\bf 125}, 217402}~(2020).

\bibitem{demleradf}
Yuto Ashida, Ata\ifmmode \mbox{\c{c}}\else~\c{c}\fi{} \ifmmode \dot{I}\else
  \.{I}\fi{}mamo\ifmmode~\breve{g}\else \u{g}\fi{}lu, and Eugene Demler.
\newblock ``Cavity quantum electrodynamics at arbitrary light-matter coupling
  strengths''.
\newblock \href{https://dx.doi.org/10.1103/PhysRevLett.126.153603}{Phys. Rev.
  Lett. {\bf 126}, 153603}~(2021).

\bibitem{mipaper2}
Beatriz P\'erez-Gonz\'alez, Miguel Bello, Gloria Platero, and \'Alvaro
  G\'omez-Le\'on.
\newblock ``Simulation of 1d topological phases in driven quantum dot arrays''.
\newblock \href{https://dx.doi.org/10.1103/PhysRevLett.123.126401}{Phys. Rev.
  Lett. {\bf 123}, 126401}~(2019).

\bibitem{AlvaroGloriaGrafeno}
Pierre Delplace, \'Alvaro G\'omez-Le\'on, and Gloria Platero.
\newblock ``Merging of dirac points and floquet topological transitions in
  ac-driven graphene''.
\newblock \href{https://dx.doi.org/10.1103/PhysRevB.88.245422}{Phys. Rev. B
  {\bf 88}, 245422}~(2013).

\bibitem{Rudner2013}
Mark~S. Rudner, Netanel~H. Lindner, Erez Berg, and Michael Levin.
\newblock ``Anomalous edge states and the bulk-edge correspondence for
  periodically driven two-dimensional systems''.
\newblock \href{https://dx.doi.org/10.1103/PhysRevX.3.031005}{Phys. Rev. X {\bf
  3}, 031005}~(2013).

\bibitem{MoraisAF}
A.~Quelle, C.~Weitenberg, K.~Sengstock, and C.~Morais~Smith.
\newblock ``Driving protocol for a floquet topological phase without static
  counterpart''.
\newblock \href{https://dx.doi.org/10.1088/1367-2630/aa8646}{New Journal of
  Physics {\bf 19}, 113010}~(2017).

\bibitem{AnomalousFI}
Frederik Nathan, Dmitry Abanin, Erez Berg, Netanel~H. Lindner, and Mark~S.
  Rudner.
\newblock ``Anomalous floquet insulators''.
\newblock \href{https://dx.doi.org/10.1103/PhysRevB.99.195133}{Phys. Rev. B
  {\bf 99}, 195133}~(2019).

\bibitem{Roy2017}
Rahul Roy and Fenner Harper.
\newblock ``Periodic table for floquet topological insulators''.
\newblock \href{https://dx.doi.org/10.1103/PhysRevB.96.155118}{Phys. Rev. B
  {\bf 96}, 155118}~(2017).

\bibitem{SSH0}
W.~P. Su, J.~R. Schrieffer, and A.~J. Heeger.
\newblock ``Solitons in polyacetylene''.
\newblock \href{https://dx.doi.org/10.1103/PhysRevLett.42.1698}{Phys. Rev.
  Lett. {\bf 42}, 1698--1701}~(1979).

\bibitem{SSH1}
A.~J. Heeger, S.~Kivelson, J.~R. Schrieffer, and W.~P. Su.
\newblock ``Solitons in conducting polymers''.
\newblock \href{https://dx.doi.org/10.1103/RevModPhys.60.781}{Rev. Mod. Phys.
  {\bf 60}, 781--850}~(1988).

\bibitem{breakdownprot}
Felice Appugliese, Josefine Enkner, Gian~Lorenzo Paravicini-Bagliani, Mattias
  Beck, Christian Reichl, Werner Wegscheider, Giacomo Scalari, Cristiano Ciuti,
  and Jérôme Faist.
\newblock ``Breakdown of topological protection by cavity vacuum fields in the
  integer quantum hall effect''.
\newblock \href{https://dx.doi.org/10.1126/science.abl5818}{Science {\bf 375},
  1030--1034}~(2022).

\bibitem{romanroche2024cavityqedmaterialscomparison}
Juan Román-Roche, Álvaro Gómez-León, Fernando Luis, and David Zueco.
\newblock ``Cavity qed materials: Comparison and validation of two linear
  response theories at arbitrary light-matter coupling strengths''~(2024).
\newblock  \href{http://arxiv.org/abs/2406.11971}{arXiv:2406.11971}.

\bibitem{gaugefixingTB}
Olesia Dmytruk and Marco Schir\'o.
\newblock ``Gauge fixing for strongly correlated electrons coupled to quantum
  light''.
\newblock \href{https://dx.doi.org/10.1103/PhysRevB.103.075131}{Phys. Rev. B
  {\bf 103}, 075131}~(2021).

\bibitem{Feynman-1951}
Richard~P. Feynman.
\newblock ``An operator calculus having applications in quantum
  electrodynamics''.
\newblock \href{https://dx.doi.org/10.1103/PhysRev.84.108}{Phys. Rev. {\bf 84},
  108--128}~(1951).

\bibitem{Mahan}
Gerald~D. Mahan.
\newblock ``Many-particle physics''.
\newblock
  \href{https://dx.doi.org/https://doi.org/10.1007/978-1-4757-5714-9}{Springer
  New York, NY}. New York~(2000).

\bibitem{TopSymmetries}
Ching-Kai Chiu, Jeffrey C.~Y. Teo, Andreas~P. Schnyder, and Shinsei Ryu.
\newblock ``Classification of topological quantum matter with symmetries''.
\newblock \href{https://dx.doi.org/10.1103/RevModPhys.88.035005}{Rev. Mod.
  Phys. {\bf 88}, 035005}~(2016).

\bibitem{classTIandSC}
Andreas~P. Schnyder, Shinsei Ryu, Akira Furusaki, and Andreas W.~W. Ludwig.
\newblock ``Classification of topological insulators and superconductors in
  three spatial dimensions''.
\newblock \href{https://dx.doi.org/10.1103/PhysRevB.78.195125}{Phys. Rev. B
  {\bf 78}, 195125}~(2008).

\bibitem{mipaper1}
Beatriz P\'erez-Gonz\'alez, Miguel Bello, \'Alvaro G\'omez-Le\'on, and Gloria
  Platero.
\newblock ``Interplay between long-range hopping and disorder in topological
  systems''.
\newblock \href{https://dx.doi.org/10.1103/PhysRevB.99.035146}{Phys. Rev. B
  {\bf 99}, 035146}~(2019).

\bibitem{manipulatingTruncation}
Martin Kiffner, Jonathan~R. Coulthard, Frank Schlawin, Arzhang Ardavan, and
  Dieter Jaksch.
\newblock ``Manipulating quantum materials with quantum light''.
\newblock \href{https://dx.doi.org/10.1103/PhysRevB.99.085116}{Phys. Rev. B
  {\bf 99}, 085116}~(2019).

\bibitem{mottpolaritons}
Martin Kiffner, Jonathan Coulthard, Frank Schlawin, Arzhang Ardavan, and Dieter
  Jaksch.
\newblock ``Mott polaritons in cavity-coupled quantum materials''.
\newblock \href{https://dx.doi.org/10.1088/1367-2630/ab31c7}{New Journal of
  Physics {\bf 21}, 073066}~(2019).

\bibitem{Asboth}
J.~K. Asbóth, L.~Oroszlány, and A.~Pályi.
\newblock ``A short course on topological insulators''.
\newblock
  \href{https://dx.doi.org/https://doi.org/10.1007/978-3-319-25607-8}{Springer
  Cham}. New York~(2016).

\bibitem{bulkboundary}
Yasuhiro Hatsugai.
\newblock ``Chern number and edge states in the integer quantum hall effect''.
\newblock \href{https://dx.doi.org/10.1103/PhysRevLett.71.3697}{Phys. Rev.
  Lett. {\bf 71}, 3697--3700}~(1993).

\bibitem{Gurarie2011}
V.~Gurarie.
\newblock ``Single-particle green's functions and interacting topological
  insulators''.
\newblock \href{https://dx.doi.org/10.1103/PhysRevB.83.085426}{Phys. Rev. B
  {\bf 83}, 085426}~(2011).

\bibitem{Gurarie2011-2}
Andrew~M. Essin and Victor Gurarie.
\newblock ``Bulk-boundary correspondence of topological insulators from their
  respective green's functions''.
\newblock \href{https://dx.doi.org/10.1103/PhysRevB.84.125132}{Phys. Rev. B
  {\bf 84}, 125132}~(2011).

\bibitem{Gurarie2012}
Salvatore~R. Manmana, Andrew~M. Essin, Reinhard~M. Noack, and Victor Gurarie.
\newblock ``Topological invariants and interacting one-dimensional fermionic
  systems''.
\newblock \href{https://dx.doi.org/10.1103/PhysRevB.86.205119}{Phys. Rev. B
  {\bf 86}, 205119}~(2012).

\bibitem{Shiozaki2018}
Ken Shiozaki, Hassan Shapourian, Kiyonori Gomi, and Shinsei Ryu.
\newblock ``Many-body topological invariants for fermionic short-range
  entangled topological phases protected by antiunitary symmetries''.
\newblock \href{https://dx.doi.org/10.1103/PhysRevB.98.035151}{Phys. Rev. B
  {\bf 98}, 035151}~(2018).

\bibitem{Shapourian2017}
Hassan Shapourian, Ken Shiozaki, and Shinsei Ryu.
\newblock ``Many-body topological invariants for fermionic symmetry-protected
  topological phases''.
\newblock \href{https://dx.doi.org/10.1103/PhysRevLett.118.216402}{Phys. Rev.
  Lett. {\bf 118}, 216402}~(2017).

\bibitem{nguyen2024electronconductancecavityembeddedtopological}
Danh-Phuong Nguyen, Geva Arwas, and Cristiano Ciuti.
\newblock ``Electron conductance and many-body marker of a cavity-embedded
  topological one-dimensional chain''.
\newblock \href{https://dx.doi.org/10.1103/PhysRevB.110.195416}{Phys. Rev. B
  {\bf 110}, 195416}~(2024).

\bibitem{gomezleon2024highqualitypoormansmajorana}
Álvaro Gómez-León, Marco Schirò, and Olesia Dmytruk.
\newblock ``High-quality poor man's majorana bound states from cavity
  embedding''~(2024).
\newblock  \href{http://arxiv.org/abs/2407.12088}{arXiv:2407.12088}.

\bibitem{relevanceP2}
Christian Sch{\"a}fer, Michael Ruggenthaler, Vasil Rokaj, and Angel Rubio.
\newblock ``Relevance of the quadratic diamagnetic and self-polarization terms
  in cavity quantum electrodynamics''.
\newblock \href{https://dx.doi.org/10.1021/acsphotonics.9b01649}{ACS Photonics
  {\bf 7}, 975--990}~(2020).

\bibitem{Hepp1973}
Klaus Hepp and Elliott~H. Lieb.
\newblock ``Equilibrium statistical mechanics of matter interacting with the
  quantized radiation field''.
\newblock \href{https://dx.doi.org/10.1103/PhysRevA.8.2517}{Phys. Rev. A {\bf
  8}, 2517--2525}~(1973).

\bibitem{Wang1973}
Y.~K. Wang and F.~T. Hioe.
\newblock ``Phase transition in the dicke model of superradiance''.
\newblock \href{https://dx.doi.org/10.1103/PhysRevA.7.831}{Phys. Rev. A {\bf
  7}, 831--836}~(1973).

\bibitem{RomanZueco2022}
Juan Rom\'an-Roche and David Zueco.
\newblock ``{Effective theory for matter in non-perturbative cavity QED}''.
\newblock \href{https://dx.doi.org/10.21468/SciPostPhysLectNotes.50}{SciPost
  Phys. Lect. NotesPage~50}~(2022).

\bibitem{romanroche2024linearresponsetheorycavity}
Juan Román-Roche, Álvaro Gómez-León, Fernando Luis, and David Zueco.
\newblock ``Linear response theory for cavity qed materials''~(2024).
\newblock  \href{http://arxiv.org/abs/2406.11957}{arXiv:2406.11957}.

\bibitem{nogoTruncation}
G.~M. Andolina, F.~M.~D. Pellegrino, V.~Giovannetti, A.~H. MacDonald, and
  M.~Polini.
\newblock ``Cavity quantum electrodynamics of strongly correlated electron
  systems: A no-go theorem for photon condensation''.
\newblock \href{https://dx.doi.org/10.1103/PhysRevB.100.121109}{Phys. Rev. B
  {\bf 100}, 121109}~(2019).

\bibitem{Blais2004}
Alexandre Blais, Ren-Shou Huang, Andreas Wallraff, S.~M. Girvin, and R.~J.
  Schoelkopf.
\newblock ``Cavity quantum electrodynamics for superconducting electrical
  circuits: An architecture for quantum computation''.
\newblock \href{https://dx.doi.org/10.1103/PhysRevA.69.062320}{Phys. Rev. A
  {\bf 69}, 062320}~(2004).

\bibitem{Lupascu2007}
A.~Lupa{\c{s}}cu, S.~Saito, T.~Picot, P.~C. de~Groot, C.~J. P.~M. Harmans, and
  J.~E. Mooij.
\newblock ``Quantum non-demolition measurement of a superconducting two-level
  system''.
\newblock \href{https://dx.doi.org/10.1038/nphys509}{Nature Physics {\bf 3},
  119--123}~(2007).

\bibitem{Nakajima2019}
Takashi Nakajima, Akito Noiri, Jun Yoneda, Matthieu~R. Delbecq, Peter Stano,
  Tomohiro Otsuka, Kenta Takeda, Shinichi Amaha, Giles Allison, Kento Kawasaki,
  Arne Ludwig, Andreas~D. Wieck, Daniel Loss, and Seigo Tarucha.
\newblock ``Quantum non-demolition measurement of an electron spin qubit''.
\newblock \href{https://dx.doi.org/10.1038/s41565-019-0426-x}{Nature
  Nanotechnology {\bf 14}, 555--560}~(2019).

\bibitem{GomezLeon2022}
\'Alvaro G\'omez-Le\'on, Fernando Luis, and David Zueco.
\newblock ``Dispersive readout of molecular spin qudits''.
\newblock \href{https://dx.doi.org/10.1103/PhysRevApplied.17.064030}{Phys. Rev.
  Applied {\bf 17}, 064030}~(2022).

\bibitem{TopologyDet}
Beatriz P\'erez-Gonz\'alez, \'Alvaro G\'omez-Le\'on, and Gloria Platero.
\newblock ``Topology detection in cavity qed''.
\newblock \href{https://dx.doi.org/10.1039/D2CP01806C}{Phys. Chem. Chem. Phys.
  {\bf 24}, 15860--15870}~(2022).

\bibitem{Leijnse2011}
Martin Leijnse and Karsten Flensberg.
\newblock ``Quantum information transfer between topological and spin qubit
  systems''.
\newblock \href{https://dx.doi.org/10.1103/PhysRevLett.107.210502}{Phys. Rev.
  Lett. {\bf 107}, 210502}~(2011).

\bibitem{Lang2017}
Lang Nicolai and Hans~Peter Büchler.
\newblock ``Topological networks for quantum communication between distant
  qubits''.
\newblock \href{https://dx.doi.org/10.1038/s41534-017-0047-x}{npj Quantum
  Information {\bf 3}, 47}~(2017).

\bibitem{Miguel2017}
M.~Bello, C.~E. Creffield, and G.~Platero.
\newblock ``Sublattice dynamics and quantum state transfer of doublons in
  two-dimensional lattices''.
\newblock \href{https://dx.doi.org/10.1103/PhysRevB.95.094303}{Phys. Rev. B
  {\bf 95}, 094303}~(2017).

\bibitem{PhysRevB.84.195413}
O.~V. Kibis, O.~Kyriienko, and I.~A. Shelykh.
\newblock ``Band gap in graphene induced by vacuum fluctuations''.
\newblock \href{https://dx.doi.org/10.1103/PhysRevB.84.195413}{Phys. Rev. B
  {\bf 84}, 195413}~(2011).

\bibitem{PhysRevB.99.235156}
Xiao Wang, Enrico Ronca, and Michael~A. Sentef.
\newblock ``Cavity quantum electrodynamical chern insulator: Towards
  light-induced quantized anomalous hall effect in graphene''.
\newblock \href{https://dx.doi.org/10.1103/PhysRevB.99.235156}{Phys. Rev. B
  {\bf 99}, 235156}~(2019).

\bibitem{PhysRevLett.118.126803}
Matthieu~C. Dartiailh, Takis Kontos, Benoit Dou\ifmmode~\mbox{\c{c}}\else
  \c{c}\fi{}ot, and Audrey Cottet.
\newblock ``Direct cavity detection of majorana pairs''.
\newblock \href{https://dx.doi.org/10.1103/PhysRevLett.118.126803}{Phys. Rev.
  Lett. {\bf 118}, 126803}~(2017).

\bibitem{PhysRevB.107.115418}
Olesia Dmytruk and Mircea Trif.
\newblock ``Microwave detection of gliding majorana zero modes in nanowires''.
\newblock \href{https://dx.doi.org/10.1103/PhysRevB.107.115418}{Phys. Rev. B
  {\bf 107}, 115418}~(2023).

\bibitem{PhysRevLett.109.257002}
Mircea Trif and Yaroslav Tserkovnyak.
\newblock ``Resonantly tunable majorana polariton in a microwave cavity''.
\newblock \href{https://dx.doi.org/10.1103/PhysRevLett.109.257002}{Phys. Rev.
  Lett. {\bf 109}, 257002}~(2012).

\bibitem{ultrastronglySCQ}
Shuai-Peng Wang, Guo-Qiang Zhang, Yimin Wang, Zhen Chen, Tiefu Li, J.~S. Tsai,
  Shi-Yao Zhu, and J.~Q. You.
\newblock ``Photon-dressed bloch-siegert shift in an ultrastrongly coupled
  circuit quantum electrodynamical system''.
\newblock \href{https://dx.doi.org/10.1103/PhysRevApplied.13.054063}{Phys. Rev.
  Appl. {\bf 13}, 054063}~(2020).

\bibitem{uscscq}
P.~Forn-D\'{\i}az, J.~Lisenfeld, D.~Marcos, J.~J. Garc\'{\i}a-Ripoll,
  E.~Solano, C.~J. P.~M. Harmans, and J.~E. Mooij.
\newblock ``Observation of the bloch-siegert shift in a qubit-oscillator system
  in the ultrastrong coupling regime''.
\newblock \href{https://dx.doi.org/10.1103/PhysRevLett.105.237001}{Phys. Rev.
  Lett. {\bf 105}, 237001}~(2010).

\bibitem{uscscq2}
T.~Niemczyk, F.~Deppe, H.~Huebl, E.~P. Menzel, F.~Hocke, M.~J. Schwarz, J.~J.
  Garcia-Ripoll, D.~Zueco, T.~H{\"u}mmer, E.~Solano, A.~Marx, and R.~Gross.
\newblock ``Circuit quantum electrodynamics in the ultrastrong-coupling
  regime''.
\newblock \href{https://dx.doi.org/10.1038/nphys1730}{Nature Physics {\bf 6},
  772--776}~(2010).

\bibitem{expsshsq}
W.~Cai, J.~Han, Feng Mei, Y.~Xu, Y.~Ma, X.~Li, H.~Wang, Y.~P. Song, Zheng-Yuan
  Xue, Zhang-qi Yin, Suotang Jia, and Luyan Sun.
\newblock ``Observation of topological magnon insulator states in a
  superconducting circuit''.
\newblock \href{https://dx.doi.org/10.1103/PhysRevLett.123.080501}{Phys. Rev.
  Lett. {\bf 123}, 080501}~(2019).

\bibitem{sshLCcircuit}
Tal Goren, Kirill Plekhanov, F\'elicien Appas, and Karyn Le~Hur.
\newblock ``Topological zak phase in strongly coupled lc circuits''.
\newblock \href{https://dx.doi.org/10.1103/PhysRevB.97.041106}{Phys. Rev. B
  {\bf 97}, 041106}~(2018).

\bibitem{2Dsshscq}
Lu~Qi, Qiao-Nan Li, Xu-Hui Yan, Wei Du, Xiuyun Zhang, and Ai-Lei He.
\newblock ``Disorder-enhanced corner state and topological corner state
  transfer in a 2d rice–mele lattice''.
\newblock
  \href{https://dx.doi.org/https://doi.org/10.1002/qute.202300369}{Advanced
  Quantum Technologies {\bf 7}, 2300369}~(2024).

\bibitem{Maschler2008}
C.~Maschler, I.~B. Mekhov, and H.~Ritsch.
\newblock ``Ultracold atoms in optical lattices generated by quantized
  light fields''.
\newblock \href{https://dx.doi.org/10.1140/epjd/e2008-00016-4}{The European
  Physical Journal D {\bf 46}, 545--560}~(2008).

\bibitem{sshcoldatoms}
Marcos Atala, Monika Aidelsburger, Julio~T. Barreiro, Dmitry Abanin, Takuya
  Kitagawa, Eugene Demler, and Immanuel Bloch.
\newblock ``Direct measurement of the zak phase in topological bloch bands''.
\newblock \href{https://dx.doi.org/10.1038/nphys2790}{Nature Physics {\bf 9},
  795--800}~(2013).

\bibitem{Meier2016}
Eric~J. Meier, Fangzhao~Alex An, and Bryce Gadway.
\newblock ``Observation of the topological soliton state in the
  su--schrieffer--heeger model''.
\newblock \href{https://dx.doi.org/10.1038/ncomms13986}{Nature Communications
  {\bf 7}, 13986}~(2016).

\bibitem{sshRydberg}
Sylvain de~Léséleuc, Vincent Lienhard, Pascal Scholl, Daniel Barredo,
  Sebastian Weber, Nicolai Lang, Hans~Peter Büchler, Thierry Lahaye, and
  Antoine Browaeys.
\newblock ``Observation of a symmetry-protected topological phase of
  interacting bosons with rydberg atoms''.
\newblock \href{https://dx.doi.org/10.1126/science.aav9105}{Science {\bf 365},
  775--780}~(2019).

\bibitem{coldatomlatticecavityqed}
A.~Pi\~neiro Orioli, J.~K. Thompson, and A.~M. Rey.
\newblock ``Emergent dark states from superradiant dynamics in multilevel atoms
  in a cavity''.
\newblock \href{https://dx.doi.org/10.1103/PhysRevX.12.011054}{Phys. Rev. X
  {\bf 12}, 011054}~(2022).

\bibitem{topinsincoldatomcavity}
Titas Chanda, Rebecca Kraus, Giovanna Morigi, and Jakub Zakrzewski.
\newblock ``Self-organized topological insulator due to cavity-mediated
  correlated tunneling''.
\newblock \href{https://dx.doi.org/10.22331/q-2021-07-13-501}{{Quantum} {\bf
  5}, 501}~(2021).

\bibitem{boseglassphasescavityqed}
Hessam Habibian, Andr\'e Winter, Simone Paganelli, Heiko Rieger, and Giovanna
  Morigi.
\newblock ``Bose-glass phases of ultracold atoms due to cavity backaction''.
\newblock \href{https://dx.doi.org/10.1103/PhysRevLett.110.075304}{Phys. Rev.
  Lett. {\bf 110}, 075304}~(2013).

\bibitem{manybodyquantumlight}
T.~J. Elliott and I.~B. Mekhov.
\newblock ``Engineering many-body dynamics with quantum light potentials and
  measurements''.
\newblock \href{https://dx.doi.org/10.1103/PhysRevA.94.013614}{Phys. Rev. A
  {\bf 94}, 013614}~(2016).

\bibitem{coldatomlatticecav}
Catalin-Mihai Halati, Ameneh Sheikhan, Helmut Ritsch, and Corinna Kollath.
\newblock ``Numerically exact treatment of many-body self-organization in a
  cavity''.
\newblock \href{https://dx.doi.org/10.1103/PhysRevLett.125.093604}{Phys. Rev.
  Lett. {\bf 125}, 093604}~(2020).

\bibitem{longrangequasipqed}
Piotr Kubala, Piotr Sierant, Giovanna Morigi, and Jakub Zakrzewski.
\newblock ``Ergodicity breaking with long-range cavity-induced quasiperiodic
  interactions''.
\newblock \href{https://dx.doi.org/10.1103/PhysRevB.103.174208}{Phys. Rev. B
  {\bf 103}, 174208}~(2021).

\bibitem{coldatomcavitypot}
Christoph Maschler and Helmut Ritsch.
\newblock ``Cold atom dynamics in a quantum optical lattice potential''.
\newblock \href{https://dx.doi.org/10.1103/PhysRevLett.95.260401}{Phys. Rev.
  Lett. {\bf 95}, 260401}~(2005).

\bibitem{higherorderPeierls}
Joana Fraxanet, Alexandre Dauphin, Maciej Lewenstein, Luca Barbiero, and Daniel
  Gonz\'alez-Cuadra.
\newblock ``Higher-order topological peierls insulator in a two-dimensional
  atom-cavity system''.
\newblock \href{https://dx.doi.org/10.1103/PhysRevLett.131.263001}{Phys. Rev.
  Lett. {\bf 131}, 263001}~(2023).

\bibitem{cavitybackaction}
Hessam Habibian, Andr\'e Winter, Simone Paganelli, Heiko Rieger, and Giovanna
  Morigi.
\newblock ``Bose-glass phases of ultracold atoms due to cavity backaction''.
\newblock \href{https://dx.doi.org/10.1103/PhysRevLett.110.075304}{Phys. Rev.
  Lett. {\bf 110}, 075304}~(2013).

\bibitem{Landig2016}
Renate Landig, Lorenz Hruby, Nishant Dogra, Manuele Landini, Rafael Mottl,
  Tobias Donner, and Tilman Esslinger.
\newblock ``Quantum phases from competing short- and long-range interactions in
  an optical lattice''.
\newblock \href{https://dx.doi.org/10.1038/nature17409}{Nature {\bf 532},
  476--479}~(2016).

\bibitem{mottinsuldickehbb}
J.~Klinder, H.~Ke\ss{}ler, M.~Reza Bakhtiari, M.~Thorwart, and A.~Hemmerich.
\newblock ``Observation of a superradiant mott insulator in the dicke-hubbard
  model''.
\newblock \href{https://dx.doi.org/10.1103/PhysRevLett.115.230403}{Phys. Rev.
  Lett. {\bf 115}, 230403}~(2015).

\bibitem{FriskKockum2019}
Anton Frisk~Kockum, Adam Miranowicz, Simone De~Liberato, Salvatore Savasta, and
  Franco Nori.
\newblock ``Ultrastrong coupling between light and matter''.
\newblock \href{https://dx.doi.org/10.1038/s42254-018-0006-2}{Nature Reviews
  Physics {\bf 1}, 19--40}~(2019).

\bibitem{QEDarchQC}
Alexandre Blais, Ren-Shou Huang, Andreas Wallraff, S.~M. Girvin, and R.~J.
  Schoelkopf.
\newblock ``Cavity quantum electrodynamics for superconducting electrical
  circuits: An architecture for quantum computation''.
\newblock \href{https://dx.doi.org/10.1103/PhysRevA.69.062320}{Phys. Rev. A
  {\bf 69}, 062320}~(2004).

\bibitem{approaching1}
A.~Wallraff, D.~I. Schuster, A.~Blais, L.~Frunzio, J.~Majer, M.~H. Devoret,
  S.~M. Girvin, and R.~J. Schoelkopf.
\newblock ``Approaching unit visibility for control of a superconducting qubit
  with dispersive readout''.
\newblock \href{https://dx.doi.org/10.1103/PhysRevLett.95.060501}{Phys. Rev.
  Lett. {\bf 95}, 060501}~(2005).

\bibitem{qndDispersive}
Maxime Boissonneault, J.~M. Gambetta, and Alexandre Blais.
\newblock ``Dispersive regime of circuit qed: Photon-dependent qubit dephasing
  and relaxation rates''.
\newblock \href{https://dx.doi.org/10.1103/PhysRevA.79.013819}{Phys. Rev. A
  {\bf 79}, 013819}~(2009).

\bibitem{qndDispersive2}
Jay Gambetta, Alexandre Blais, D.~I. Schuster, A.~Wallraff, L.~Frunzio,
  J.~Majer, M.~H. Devoret, S.~M. Girvin, and R.~J. Schoelkopf.
\newblock ``Qubit-photon interactions in a cavity: Measurement-induced
  dephasing and number splitting''.
\newblock \href{https://dx.doi.org/10.1103/PhysRevA.74.042318}{Phys. Rev. A
  {\bf 74}, 042318}~(2006).

\end{thebibliography}

\appendix

\section{LIGHT-MATTER COUPLING IN THE COULOMB AND DIPOLE GAUGE \label{ap:coulomb}}

The starting point is the tight-binding Hamiltonian $H_{\mathrm{el}}=\sum_{ij}^{N}J_{i,j}c_{i}^{\dagger}c_{j}$ that is coupled to a single-mode photonic field $H_{\mathrm{phot}} = \Omega d^\dagger d$. In general, the minimal-coupling Hamiltonian describing the light-matter interaction in the Coulomb gauge can be implemented through an unitary transformation $U$ acting on the fermionic degrees of freedom only \cite{resolutionNori, gaugefixingTB, giDickHopfield}, with

\begin{equation}
	U = \mathrm{exp}\left\{i (d^\dagger + d) \sum_{ij} \chi_{ij} c^\dagger_i c_j \right\}
	\label{eq:transf}
\end{equation}

where $\chi_{ij} = \langle i| \chi |j \rangle = \int d\vec{r} \phi^*_i(\vec{r}) \chi(\vec{r}) \phi_j(\vec{r})$, being $\phi_i(\vec{r})$ the single-particle wavefunctions localized around each lattice site $i$, and $\chi(\vec{r})$ chosen such that $\nabla \chi(\vec{r}) = e A_0(\vec{r})$, or equivalently,

\begin{equation}
	\chi(\vec{r}) = e \int_{\vec{r}_0}^{\vec{r}} A_0 (\vec{r}^\prime)\cdot d\vec{r}^\prime
	\label{eq:chi}
\end{equation}

In general, the vector potential in a 1D system can be written as $\vec{A}(r) = A_0 (r)(d^\dagger + d)\hat{u}(r)$, where $A_0 = 1/\sqrt{2\varepsilon_0 \Omega \nu}$ $(\hbar = 1)$, with $\varepsilon_0$ being the vacuum permittivity and $\nu$ the cavity volume. Under the dipole approximation used in the main text $A_0(r) \approx A_0$, and assuming that the vector potential is orientated along the axis of the electronic chain (without loss of generality, we will assume this is the $x$ axis), this expression can be further simplified to yield $\chi(r) = e A_0 r$, where we have dropped the vectorial notation for the position coordinate (we have also taken $r_0 = 0$). Therefore, we can also write $\chi_{ij} \approx \delta_{ij} \chi_{ii} = e A_0 r_i \delta_{ij}$, and the final form of the transformation $U$ is

\begin{equation}
	U = \mathrm{exp}\left\{i e A_0 (d^\dagger + d) \sum_{i} r_i c^\dagger_i c_i \right\}
\end{equation}

Using the Baker-Campbell-Hausdorff formula, the transformed fermionic operators will then give

\begin{equation}
	U^\dagger c_i U = c_i e^{i e A_0 r_i (d^\dagger + d)},\hspace{3pt} U^\dagger c_i^\dagger U = c_i^\dagger e^{-i e A_0 r_i (d^\dagger + d)}
\end{equation}

The Peierls Hamiltonian $H$ shown in Eq. $(1)$ in the main text can be obtained by operating $H_{\mathrm{PS}} = H_\Omega + U^\dagger H_\mathrm{el} U$.\\

In conclusion, the Peierls substitution yields a minimal-coupling Hamiltonian under the dipole approximation, satisfying the gauge principle. Note that further orbital structure $c^{(\dagger)}_i \rightarrow c^{(\dagger)}_{i\mu}$ has not been taken into account, in which case there would be additional dipole-like terms connecting different orbitals. In general, without applying the dipole approximation and including a nontrivial structure in both real and orbital space, the transformation of the fermionic operators would give 

\begin{equation}
	U^\dagger c_i U = \mathrm{exp}\left\{i(d^\dagger + d)\sum_{ij, \mu \mu^\prime} \chi_{ij}^{\mu \mu^\prime}c^\dagger_{i\mu} c_{j\mu^\prime} \right\}
\end{equation}

The results of this section, and in particular Eqs. \ref{eq:transf} and \ref{eq:chi}, are also in agreement with the parallel transporter introduced in \cite{gaugeinvZueco} to derive gauge-invariant Hamiltonians using lattice gauge theory for two-level systems, and with the results obtained for the Dicke and Hopfield models in \cite{giDickHopfield}.\\

An alternative form of the light-matter coupling can be obtained by writting the Peierls Hamiltonian in the dipole gauge (DG). This transformation, that we will denote $T$, is the inverse of $U$

\begin{equation}
	T= U^\dagger,\hspace{5pt} T = \mathrm{exp}\left\{ - i e A_0 (d^\dagger + d)\sum_i r_i c^\dagger_i c_i \right\}
\end{equation}

and is applied only on the photonic operators, $H_{DG} = T^\dagger H_\Omega T + H_\mathrm{el}$. In particular, the transformed photonic operators give

\begin{equation}
	T^\dagger d T = d + i\sum_j r_j c^\dagger_j c_j,\hspace{5pt} T^\dagger d T = d -i\sum_i r_i c^\dagger_i c_i
\end{equation}

and $H_\mathrm{DG}$ has the following form

\begin{eqnarray}
	H_{\mathrm{DG}} & = & \Omega d^\dagger d + H_{el} + i e A_0 \Omega (d - d^\dagger )\sum_i r_i c^\dagger_i c_i \\
	& & + e A_0\Omega \left(\sum_{i} r_i c^\dagger_i c_i \right)^2 \nonumber
	\label{eq:dipolegaugeS}
\end{eqnarray}

as shown in the main text.\\

\section{DERIVATION OF THE PEIERLS HAMILTONIAN IN THE BASIS OF PHOTON STATES \label{sec:laguerre_krummer}}

In this section, we detail the derivation of the Peierls Hamiltonian in the photon basis. We first apply the Baker-Haussdorf-Campbell identity for non-commuting operators $A$ and $B$: $e^{A + B} = e^{-[A,B]/2}e^{A}e^{B}$, to the Peierls operator:
\begin{equation}
	e^{i\eta_{ji}(d^\dagger + d)} = e^{-\eta^2_{ji}/2}e^{i\eta_{ji}d^\dagger }e^{i\eta_{ji}d},
\end{equation}

This allows us to factor and re-organize the different powers of the creation and annihilation operators, which now are weighted by an additional exponential pre-factor $e^{-\eta^2_{ji}/2}$. Applied on the photonic number state $|n\rangle$, one gets

\begin{equation}
	e^{i\eta_{ji}\left(d^{\dagger}+d\right)} |n\rangle = e^{-\eta_{ji}^{2}/2}\sum_{q,k=0} \Theta(q,n,k) |n-k+q \rangle
\end{equation}

where we have defined

\begin{equation}
	\Theta(q,n,k) = \frac{\left(i\eta\right)^{q+k} \sqrt{\left(n-k+q\right)!n!}}{q!k!\left(n-k\right)!}
\end{equation}

where we have used the definitions of the creation/annihilation operators $d\vert n\rangle = \sqrt{n}\vert n - 1\rangle $ and $d^\dagger \vert n\rangle = \sqrt{n + 1}\vert n + 1\rangle$. Then, for any two given photonic subspaces $l$ and $s$, the prefactors $g^{ji}_{ls}$ are:
\begin{equation}
	g^{ji}_{ls} = \langle l | e^{i \eta_{ji}(d^\dagger + d)}|s \rangle = e^{-\eta^2_{ji}/2} \sum_{q=0}^{\infty} \sum_{k=0}^s \frac{(i \eta_{ji})^{q+k}}{q! k! (s - k)!}\sqrt{l! s!}
\end{equation}

For $l=s$ (diagonal terms), this finite sum converges to the Laguerre polynomials $L_s$, and the off-diagonal terms are proportional to the Krummer's confluent hypergeometric function $_{1}F_{1}(s,l,\eta^2)$, as shown in the main text.\\

\section{SSH HAMILTONIAN \label{ap:SSH}}

\begin{figure}
	\centering
	\includegraphics{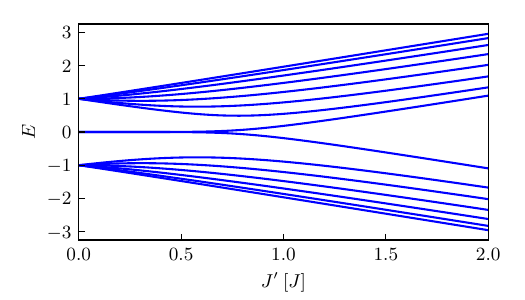}
	\caption{ \label{fig:sshspectrum} Energy spectrum of the SSH model as a function of the hopping amplitudes $J^\prime$, with $J=1$ fixed. In the left side of the plot $(J^\prime < J$), the presence of in-gap edge states indicates that the system has non-trivial topology.}
\end{figure}

The SSH model is a tight-binding model describing non-interacting electrons in a 1D lattice with an alternating pattern of first-neighbour hoppings, $J^\prime$ and $J$. This dimerized structure motivates the definition of two sublattices, $A$ and $B$, and the use of sublattice creation/annihilation operators, $a^{(\dagger)}_{\alpha}$ and $b^{(\dagger)}_{\alpha}$, where $\alpha = 1,2,.., N$ is a cell index and $N$ is the total number of cells/dimers. Its second-quantization Hamiltonian can be written as

\begin{equation}
	H_{\mathrm{SSH}} = \sum_\alpha^N  \big( J^\prime a^\dagger_\alpha b_\alpha + J b^\dagger_\alpha a_{\alpha + 1} \big) + \mathrm{h.c.}.
	\label{eq:sshHamiltonianreal}
\end{equation}

Notice that the hopping $J^\prime$ takes place within the same dimer, and that is why it is usually referred to as the intra-dimer hopping, while $J$ is the inter-dimer hopping, connecting sites belonging to different dimers. Fig. \ref{fig:sshspectrum} shows the energy spectrum as a function of the hopping amplitude $J^\prime$, for a fixed $J = 1$. For $J^\prime < J$, the presence of in-gap topological edge states denotes that the chain has non-trivial topology. 

Assuming periodic-boundary conditions, the fermionic operators can be Fourier transformed to $(a_\alpha , b_\alpha) = \frac{1}{\sqrt{N}}\sum_k (a_k e^{i k r^{A}_\alpha}, b_k e^{i k r^{B}_\alpha})$, with $r^{A/B}_\alpha$ being the position of each site of the chain. The Hamiltonian is then written in the form

\begin{equation}
	H_{\mathrm{SSH}} = \sum_k (a^\dagger_k, b_k) \mathcal{H}_{\mathrm{SSH}}(k) \left( \begin{array}{c} a_k \\ b_k \end{array} \right) ,
\end{equation}

where $\mathcal{H}_{\mathrm{SSH}}(k)$ is the kernel of the Hamiltonian, defined as

\begin{equation}
	\mathcal{H}_{\mathrm{SSH}}(k) = h(k)  a^\dagger_k b_k + h^{*}(k) b^{\dagger}_k a_k,
	\label{eq:sshHamiltoniank}
\end{equation}

with $ h(k) = t^\prime e^{i k r^\prime} + t e^{-i k r}$. We can also introduce the Pauli matrices $\{ \sigma_x, \sigma_y , \sigma_z \}$ to write $\mathcal{H}_{\mathrm{SSH}} = \vec{d}(k) \cdot \vec{\sigma} $, where $\vec{d} = (\mathrm{Re}\{h(k)\}, -\mathrm{Im}\{h(k)\}, 0)$. \\

\section{EFFECT OF HIGHER-ORDER PHOTON TRANSITIONS IN THE ENERGY SPECTRUM\label{sec:higherphotons}}
\begin{figure*}
	\centering
	\includegraphics{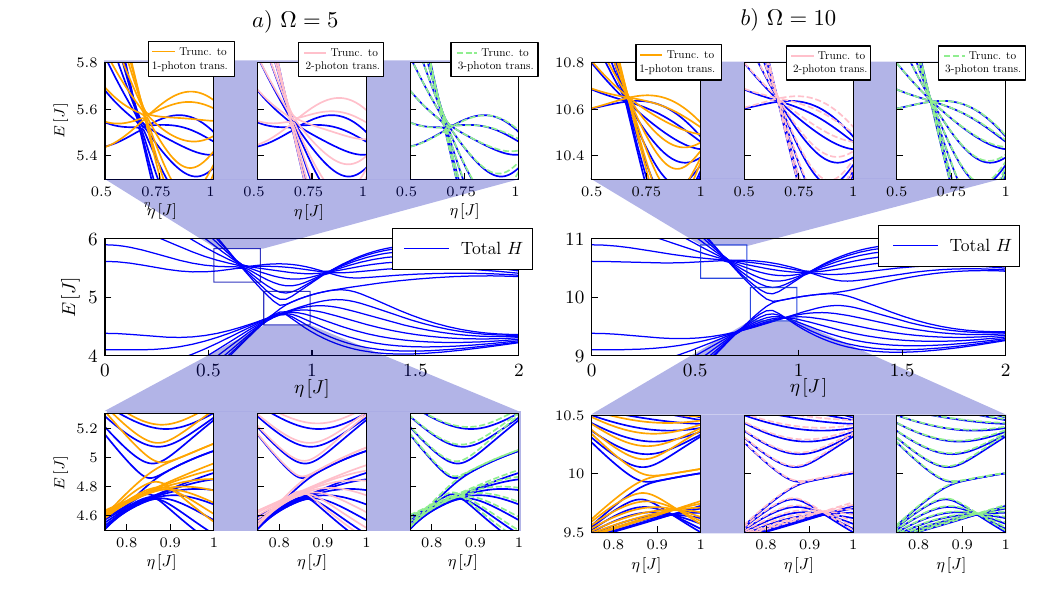}
	\caption{\label{fig:comparison} Comparison between different truncations of the total Hamiltonian (to one-, two-, and three-photon processes), as function of $\eta$ and for two cavity frequencies: $a)$ $\Omega = 5$, and $b)$ $\Omega = 10$. Parameters: $J^\prime = 1.5, $, $J = 1$, $N =8$, $n_\mathrm{max} = 50$, $r^\prime = 0.4$.}
\end{figure*}

In this section, we explore the small, qualitative mismatch between the total Hamiltonian in Eq. $(4)$ and its truncated version, originated from the effect of higher-photon processes (two-photon, three-photon transitions…). There are two parameters that have an impact on this mismatch: \\

\begin{itemize}
	\item The coupling strength: while for small values of the coupling strength the agreement between the truncated and the total Hamiltonian is perfect (see Fig. 1) in the main text, as the coupling strength increases and the system enters an ultrastrong coupling regime~\cite{FriskKockum2019}, higher-order photon transitions are expected to have a larger weight. With this, the mismatch between both Hamiltonians appears. Then, for very large values of the coupling, the two subsystems effective decouple and the differences are suppressed again, with both Hamiltonians correctly describing the system as a collection of degenerate bands separated by the cavity frequency. In fact, this is supported by Fig. $1$ in the SM, where we show the dependence of the functions $g_{m,n}$ with the coupling. Note that the off-diagonal functions $g_{m,n}$ with $m\neq n$ (controlling photon-exchange processes) are zero at $\eta= 0$ and for large enough coupling strength, while reaching its highest value for intermediate values. 
	\item The cavity frequency: For higher values of the cavity frequency, the photonic bands are further apart in energy and the exchange of photons becomes suppresses. However, as they come closer in the energy spectrum, higher-photon become more likely. This also implies that the effect of the breaking of chiral symmetry will be more evident in the energy spectrum for lower frequencies (i.e., in the assymetry of the bands). 
\end{itemize}

$a)$ the coupling strength: as the coupling strength increases, higher-order photon transitions are expected to have a larger weight,\\
$b)$ the cavity frequency: for higher values of the cavity frequency, the photonic bands are further apart in energy and the exchange of photons becomes suppresses. However, as they come closer in the energy spectrum, higher-photon transitions become more likely.\\

Fig. \ref{fig:comparison} shows the energy spectrum from the total Hamiltonian, compared to its truncation version, and how this mismatch can be overcome by including higher-order photon transitions, in particular, two- and three-photon processes. Different choices of parameters are included. In panel $a)$, $\Omega = 10$, which is the same choice as in the main text. We see how the truncated spectrums converge to the total one as more photon transitions are included. In panel $b)$, $\Omega = 5$, and the convergence is worse, making it necessary to include more photon transitions. Note that in both cases, the disagreement is just qualitative: the truncated Hamiltonian to one-photon processes is enough to capture the relevant features of the interacting system. Besides, the mismatch depends on the coupling strength as well: it is negligable for small values of $\eta^\prime$, while it becomes larger in the ultrastrong regime. \\

\section{MEAN-FIELD HAMILTONIAN \label{sec:MF}}

\begin{figure}
	\centering
	\includegraphics{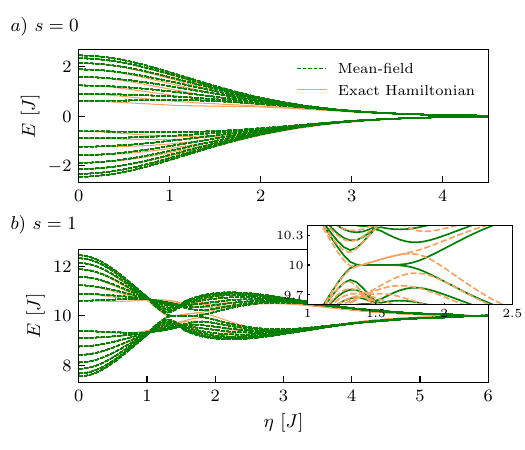}
	\caption{Comparison between the exact truncated (orange) and the mean-field fermionic Hamiltonian (green) as a function of the coupling strength $\eta^\prime$, for different photonic bands: $\mathbf{a)}$ $s=0$, and $\mathbf{b)}$ $s=1$. In $b)$, the inset shows the energy of the edge states vs. $\eta^\prime$, highlighting the difference between the chiral-symmetric MF spectrum and the truncated one in which chiral symmetry is broken. Parameters: $J^\prime = 1.5$, $J = 1$, $N = 8$ unit cells, $n_\mathrm{max} = 50$, $\Omega = 10$, $r^\prime = 0.4$.}
	\label{fig:spectrums_MF}
\end{figure}

We have shown that a simplified version of the gauge invariant Hamiltonian in the Coulomb gauge can be obtained by using disentangling techniques and truncating to one-photon processes. Now, we explore which features of the energy spectrum are retained when an even simpler approximation is used, i.e., a mean-field decoupling scheme in which each operator is rewritten as $\mathcal{O} = \langle \mathcal{O} \rangle + \delta \mathcal{O}$, separating its mean value $\langle \mathcal{O}\rangle$ from the contribution of quantum fluctuations $\delta \mathcal{O}$. At the MF level (linear terms in $\delta\mathcal{O}$), we obtain two effectively decoupled Hamiltonians for each subsystem, where the back-action due to the interaction is encoded in renormalized system parameters. An additional fluctuations Hamiltonian (second-order in $\delta\mathcal{O}$) encodes the role of quantum fluctuations, introducing correlations between subsystems. A MF analysis of the problem allows to identify which features are captured by the photo-dressing of the fermionic system and which of them are linked to electron-photon correlations. In the following, the MF approach will be used on the truncated version of the exact Hamiltonian.\\

For the photonic field, the MF decoupling on the truncated Hamiltonian gives a particularly simple description: the resulting Hamiltonian is diagonal in the basis of Hubbard operators $Y^{nn}$. First, let us define the following fermionic operators, which contain the energy contribution of both the inter- and intra-dimer hopping 

\begin{equation}
	f^{\prime}_+ = t^\prime \sum_\alpha a^\dagger_\alpha b_\alpha = (f^\prime_{-})^\dagger, \hspace{10pt} f_+ = t \sum_\alpha b^\dagger_\alpha a_{\alpha + 1} = (f_{-})^\dagger, 
	\label{eq:f}
\end{equation}
together with $ f^{(\prime)}  =  f^{(\prime)}_{+} +  f^{(\prime)}_{-}  $. Then, the MF photonic Hamiltonian can be readily written as

\begin{equation}
	H_{\text{MF,phot}} = \sum_{n} \bigg\{ n \Omega + \langle f^\prime \rangle g^{\prime}_{nn} + \langle f \rangle g_{nn}  \bigg\} Y^{nn}.
	\label{eq:diagonalHmfPhot}
\end{equation}

Note that one-photon transitions are absent in the MF photonic Hamiltonian $H_{\text{MF,phot}}$, which is diagonal in the photonic operators. This is due to the symmetry of the truncated Peierls operator in Eq. $(2)$ (main text) and the form of the off-diagonal coefficients $g^{(\prime),\pm}_{n,n+1}$. In general, the creation/annihilation of $m$ photons in the system through the operators $Y^{n+m,n}$/$Y^{n,n+m}$ is linked to the electronic inter- and intra-dimer hopping, where the corresponding tunneling amplitudes are weighted by the prefactors $g^{(\prime),\pm}_{n,n+m}$. For the case of one-photon transitions, the direction of the hopping (to the left, $\eta^{(\prime)} > 0$, or to the right, $\eta^{(\prime)} < 0$) has the effect of adding an extra sign to $g^{(\prime),\pm}_{n,n+1}$, so that they finally cancel out in $H_{\mathrm{MF,phot}}$. Note that for arbitrary $m$, all odd-photon transitions ($m=1,3,5,...$) would be absent from the MF Hamiltonian, while even ones ($m = 2,4,6,...$) are not cancelled out. This is agreement with Ref. \cite{Dmytruk2022}.\\

Then, the eigenstates of $H_{\text{MF,phot}}$ in Eq. \eqref{eq:diagonalHmfPhot} can be labelled with the photon number $n = 0,1,2,...$, being $n = 0$ the ground state with zero photons. This means that the different photonic subspaces will not hybridize for large coupling strength, and the number of photons in the system is constant in time.\\ 

The Hamiltonian in Eq. \eqref{eq:diagonalHmfPhot} indicates that the interaction with the fermionic system shifts the energy of the cavity photons, which depends on the fermionic state through $\langle f^{(\prime)}\rangle = \mathrm{Tr}\{ \rho_{\text{MF,fer}} f^{(\prime)} \}$. Being absent in the classical regime, the frequency shift is crucial in cavity QED set-ups for quantum non-demolition measurements on the fermionic system using the photonic radiated signal \cite{QEDarchQC, approaching1, qndDispersive, qndDispersive2}. \\

For the fermionic system, the MF decoupling yields an unperturbed SSH Hamiltonian with renormalized hopping amplitudes due to the interaction with the photonic field, 

\begin{equation}
	t^{(\prime)}\rightarrow \tilde{t}^{(\prime)} = t^{(\prime)}g^{(\prime)}_{s,s} = t^{(\prime)} e^{-\eta^{(\prime) 2}/2}L_s \left( \eta^{(\prime)2} \right).
	\label{eq:hopren}
\end{equation}

Note that both $\tilde{t}^{\prime}$ and $\tilde{t}$ inherit the dynamical localization prefactors that ensures the suppression of the hopping at large coupling strength, while the dependence on the Laguerre polynomial $L_s(\eta^{(\prime)2})$ gives an oscillatory dependence with $\eta^\prime$. Here we have assumed that the cavity is prepared with a fixed number of photons, $\rho_{\text{MF,phot}} = \vert s\rangle \langle s\vert$.

Fig. \ref{fig:spectrums_MF} shows the MF fermionic energy spectrum compared with the exact truncated Hamiltonian. The former reproduces the topological phase transitions as a function of the coupling strength, which indicates that topology is controlled by the dressing of the fermionic degrees of freedom (first term in Eq. $(4)$ of the main text) and the ratio of the renormalized hoppings $\tilde{t}^{\prime} / \tilde{t} $.

\section{CALCULATION OF THE TOPOLOGICAL INVARIANT FOR THE INTERACTING SYSTEM\label{sec:winding}}

An analytical solution for $G^{m,m}_{\alpha,\beta}(\omega, k)$ can be found by writing its equation of motion (EOM) in time domain

\begin{eqnarray}
	i \frac{d G^{m,m}_{\alpha,\beta}(t, k)}{dt} & = & \delta(t)\left\{ \alpha_k Y^{m,m}(t),\beta^\dagger_k \right\} \nonumber \\
	& & -i\theta(t)\langle \left\{ \left[ \alpha^{\dagger}_k Y^{m,m}, H(k) \right], \beta^\dagger_k \right\} \rangle .
	\label{eq:eom}
\end{eqnarray}

$H(k)$ is the k-space Hamiltonian of the interacting system, obtained from Fourier transforming the fermionic creation/annihilation operators into $k-$space. Its truncated version to one-photon processes is

\begin{eqnarray}
	H(k) & =	& \sum_{n}n\Omega Y^{n,n} \nonumber \\
	& & + \sum_{n}\left\{ h_{n}(k) a_{k}^{\dagger}b_{k} + h^{*}_{n}(k) b_{k}^{\dagger}a_{k}\right\} Y^{nn} \nonumber\\
	&  & + \sum_{n}\left\{ \bar{h}_{n}(k) a_{k}^{\dagger}b_{k}+\bar{h}^{*}_{n}(k)b_{k}^{\dagger}a_{k}\right\} \cdot \nonumber\\
	& & \hspace{20pt}\cdot \left(Y^{n+1,n}+Y^{n,n+1}\right)
\end{eqnarray}

together with the definitions

\begin{eqnarray}
	h_{n}(k) & = & t^{\prime}g_{nn}^{\prime}+tg_{nn}e^{ik}\\
	\bar{h}_{n}(k) & = & t^{\prime}g_{n+1,n}^{+\prime}+te^{ik}g_{n+1,n}^{-}.
\end{eqnarray}

To solve Eq. \ref{eq:eom} we have considered single occupation for the SSH chain, which greatly simplifies the fermionic algebra. Second, with the truncation of $\mathcal{H}(k)$, the EOM can be restricted to those terms involving only one-photon transitions, namely $G^{m\pm1,m}_{\alpha,\beta}(k,\omega)$ and $G^{m,m\pm 1}_{\alpha,\beta}(k,\omega)$. This entails a major reduction of the equations and enables the solution of the system, that would otherwise involve infinitely many GFs of the form $G^{m,m + l}_{\alpha,\beta}$ and $G^{m + l, m}_{\alpha,\beta}$ ($l=1,2,...$). With all this, and neglecting off-diagonal couplings, the solution for $G^{m,m}_{\alpha,\beta}(k,\omega)$ can be written in the form

\begin{equation}
	G^{m,m}(k,\omega) = \frac{\langle Y^{m,m} \rangle}{2\pi}\left[ F^{m,m}(k,\omega) \right]^{-1},
	\label{eq:Gss}
\end{equation}
where $F^{m,m}$ is a $2\times 2$ matrix which contains the pole structure for $G^{m,m}(k,\omega)$. By defining $\varepsilon^2_{m}(k) = h_{m}(k)\hspace{1pt}h^{*}_{m}(k)$ and $\bar{\varepsilon}^2_{m}(k) = \bar{h}_{m}(k)\hspace{1pt}\bar{h}^{*}_{m}(k)$, we can write the following expression for the matrix elements of $F^{m,m}$,

\begin{eqnarray}
	F^{m,m}_{A,A}(k,\omega) & = & \omega - \frac{\bar{\varepsilon}^2_{m}(k)[\omega + \Omega]}{p^{+}_{m + 1}(\omega, k)} \nonumber \\
	& & \hspace{50pt} - \frac{\bar{\varepsilon}_{m-1}(k)[\omega - \Omega]}{p^{-}_{m - 1}(\omega, k)}, \label{eq:F11}\\
	F^{m,m}_{A,B}(k,\omega) & = & - h_{m}(k) - \frac{\bar{h}_{m}^{2}(k) h_{m+1}^{*}(k)}{p^{+}_{m+1}(\omega, k)} \nonumber \\
	& & \hspace{50pt} - \frac{\bar{h}^2_{m-1}(k) h^{*}_{m-1}(k)}{p^{-}_{m-1}(\omega, k)},  \label{eq:F12} \\
	F^{m,m}_{B,A}(k,\omega) & = & \left[F^{m,m}_{A,B}(k,\omega)\right]^{*}, \nonumber \\
	& & \hspace{10pt} F_{B,B}^{m,m}(k,\omega) = F_{A,A}^{m,m}(k,\omega), \label{eq:F21yF22}
\end{eqnarray}
with $p^{+}_{m}(k,\omega) = (\omega + \Omega)^2 - \varepsilon^2_{m}(k)$ and $p^{-}_{s}(k,\omega) = (\omega - \Omega)^2 - \varepsilon^2_{m}(k)$.\\

The first term in Eqs. \eqref{eq:F11} and \eqref{eq:F12} accounts for the mean-field result of dressed electrons, while the second and third terms contain the contribution of one-photon transitions, which allows us to interprete the implications for the topology of fermionic chain.  Note that those terms contain additional poles with mixed light-matter resonances, due to the presence of correlations between subsystems.\\

Physically, the additional terms in $F^{m,m}_{\alpha,\alpha}(k,\omega)$ indicate the transfer of spectral weight to the subspaces of $(m+1)-$ (second term in Eq. \eqref{eq:F11} and $(m-1)-$photons (third term in Eq. \eqref{eq:F11}. Interestingly, since $\varepsilon^2_{m}(k) \neq \varepsilon^2_{m-1}(k)$, the transfer of spectral weight is not symmetric. Therefore, the exchange of one-photon from a given photonic band with the upper/lower bands are not equivalent.\\

\end{document}